\documentclass[aps,pra,superscriptaddress,twocolumn,showpacs,showkeys]{revtex4-2}

\usepackage{graphicx}
\usepackage{dcolumn}
\usepackage{amsmath}
\usepackage{dsfont}
\usepackage{amssymb}
\usepackage{braket}
\usepackage{bm}
\usepackage[hidelinks]{hyperref}
\usepackage{soul}
\usepackage{todonotes}
\renewcommand{\vec}[1]{\boldsymbol{#1}}
\usepackage{tcolorbox}

\usepackage{tcolorbox}
\tcbuselibrary{breakable,skins}

\newtcolorbox{highlight}{
    colback=yellow!100,
    colframe=yellow!100,
    boxrule=0pt,
    arc=0pt,
    boxsep=0pt,
    left=0pt,
    right=0pt,
    top=0pt,
    bottom=0pt,
    parbox=false,
    breakable,
    before skip=0pt,
    after skip=0pt
}

\begin{document}


\title{Strong coupling of virtual negative states in the Kapitza-Dirac effect}

\author{Qianlong Wang}
\author{Sven Ahrens}
\thanks{ahrens@shnu.edu.cn}%
\author{Baifei Shen}%
\email{bfshen@shnu.edu.cn}
\affiliation{Department of Physics, Shanghai Normal University, Shanghai 200234, China}%




\date{\today}

\begin{abstract}
Negative states are an intrinsic property of relativistic quantum theory and related to anti-particles in the context of the Dirac sea concept. We show that negative states can dominantly contribute to the diffraction amplitude in the quantum dynamics of the two-photon Kapitza-Dirac effect. We draw our conclusion by investigating solutions from time-dependent perturbation theory, where the perturbative solutions are in match with numeric solutions of the relativistic quantum system and also with the numeric and analytic solutions from the relativistic equations of motion of a classical point-like electron in an external standing wave light field. While our numeric solutions assume a strong laser field, the analytic solutions indicate that negative state coupling remains dominant for arbitrary low field amplitudes, where in the single-photon case (Compton scattering) negative state coupling can be mathematically associated with the interaction of a virtual electron-positron pair in the context of a quantized theory in old-fashioned perturbation theory.
\end{abstract}



\maketitle

\section{Introduction}
Negative solutions arise naturally in relativistic single particle quantum theory, due to quadratic energy and momentum components in the relativistic energy-momentum relation and thus imply corresponding positive and negative energy eigenstates in the quantum system. Accordingly, the Klein-Gordon and Dirac equations are known to have solutions with negative energy eigenvalues \cite{wachter_2011_relativistic_quantum_mechanics_2011,schwabl_2000_advanced_quantum_mechanics,greiner_2000_relativistic_quantum_mechanics,klein_1926_quantentheorie,gordan_1928_die_energieniveaus,dirac_1928_the_quantum_theory_of_the_electron}, which gave a theoretical indication for the existence of positrons \cite{dirac_1930_a_theory_of_electrons_and_protons,carl_1932_the_apparent_existence_of_easily_deflectable_positives}. This paradoxical situation is handled by adding up negative solutions with their positive absolute value of the energy eigenvalue in the Hamiltonian, when quantizing the fields \cite{Greiner_Field_quantization,Greiner_1985_strong_field_QED,Peskin_Schroeder_1995_Quantum_Field_Theory}. Nevertheless, it is possible to associate negative states of a single-particle theory with virtual excitations of pairs (of particles and anti-particles) in the context of such a quantized interacting theory, for the case of Compton scattering \cite{ahrens_2020_two_photon_bragg_scattering}.

An interesting aspect in this context is the circumstance that the lower two components of the Dirac equation are reinserted into the upper two components, when deducing the non-relativistic limit of the Dirac equation \cite{wachter_2011_relativistic_quantum_mechanics_2011,schwabl_2000_advanced_quantum_mechanics,greiner_2000_relativistic_quantum_mechanics}. The result of this non-relativistic limit procedure yields the well-known $(\vec p - e \vec A/c)^2$ expression in the Schr\"{o}dinger equation, which is the reason for us to provide a summary of its commonly known derivation in appendix \ref{appendix_A}. The mentioned upper and lower components, in turn, can be approximately associated with the positive and negative bi-spinor solutions of the Dirac equation for the case of non-relativistic momenta $|\vec p| \ll mc$. Therefore the question arises: How strong are those negative states actually coupled in the quantum dynamical evolution of an interacting theory with non-vanishing external fields $\vec A \neq 0$?

In this article, we investigate the coupling of negative states for the case of the Kapitza-Dirac effect \cite{kapitza_dirac_1933_proposal,batelaan_2000_KDE_first,batelaan_2007_RMP_KDE}, in which the electron is interacting with a standing wave of light. In the original proposal 1933 by Kapitza and Dirac, this effect is treated within a non-relativistic framework, utilizing the Thomson differential cross section formula \cite{jackson_1999_classical_electro_dyamics} without reference to the Dirac equation, negative-energy states, or virtual electron-positron pairs. Subsequent relativistic treatments have established the theoretical foundation for spin-dependent dynamics and high-intensity regimes \cite{ahrens_bauke_2012_spin-kde,ahrens_bauke_2013_relativistic_KDE,erhard_bauke_2015_spin,McGregor_Batelaan_2015_two_color_spin,dellweg_awwad_mueller_2016_spin-dynamics_bichromatic_laser_fields,dellweg_mueller_2016_interferometric_spin-polarizer,ahrens_2017_spin_filter,ahrens_2020_two_photon_bragg_scattering}. Building upon this foundation, our work focuses specifically on quantifying the coupling strength to negative-energy intermediate states and elucidating their role as virtual particle-antiparticle excitations in the diffraction process. A common method \cite{ahrens_bauke_2012_spin-kde,ahrens_bauke_2013_relativistic_KDE,ahrens_2017_spin_filter,ahrens_2020_two_photon_bragg_scattering} for numerically and analytically describing the quantum dynamics in terms of the energy-eigenstates allows a precise tracking of the coupling strength of negative states in the framework of time-dependent perturbation theory, which is valid in the weak field case ($|e\vec A_0| \ll mc^2$). We apply time-ordered perturbation theory techniques, as developed in quantum field theory \cite{Halzen_Martin_1984_quarks_and_leptons,Weinberg_1995_quantum_theory_of_fields}, to analyze the specific setup of the standing-wave Kapitza-Dirac effect. This approach enables a decomposition of scattering amplitudes into contributions from virtual positive- and negative-energy intermediate states and provides insight into the relativistic quantum structure of the diffraction process. The circumstance that (i) a precise experimental implementation of the Kapitza-Dirac effect already exists \cite{Freimund_Batelaan_2001_KDE_first,Freimund_Batelaan_2002_KDE_detection_PRL}, and that (ii) the mentioned association of negative states with virtual anti-particle excitations \cite{ahrens_2020_two_photon_bragg_scattering} is matching the Kapitza-Dirac effect, emphasizes the significance of this study.

Our article is structured as follows:
The relativistic quantum framework is introduced in Section \ref{sec:energy_eigenstate_dynamics}, where we first introduce the optical standing light wave in section \ref{sec:laser_set_up}. We then introduce the equations of motion (Dirac equation) for the relativistic quantum dynamics of the electron in section \ref{sec:dirac_equation}, which are subsequently solved numerically in section \ref{sec:numerical_solution_dirac_equation} for the 2-photon Kapitza-Dirac effect. A perturbative analysis of the process follows in section \ref{sec:perturbative_solution_two_kde_dirac}, which allows us to study the coupling to negative states in Section \ref{sec:negative_state_investigation}. We conclude the investigation of the relativistic electron quantum dynamics by performing an approximation of the perturbative solution for small transverse electron momenta in section \ref{sec:low_transverse_momentum_approximation}. Transitioning to the non-relativistic quantum description, Section \ref{sec:non_relativistic_quantum_description} discusses the Schr\"{o}dinger equation for modeling the electron quantum dynamics (Section \ref{sec:schrodinger_equation}), with corresponding perturbative solutions derived in Section \ref{sec:perturbative_solution_s}. Section \ref{procedure_poderomotive_tkde} introduces a non-relativistic classical treatment by deriving the ponderomotive potential, where the classical interaction with the field can be linked to the quantum processes. In Section \ref{procedure_relativistic_poderomotive_tkde} we extend this framework towards the classical relativistic equations of motion and demonstrate a match with the relativistic quantum solution. We finally discuss the role of negative states in the quantum dynamics and their relation to the classical motion in section \ref{sec:discussion} and conclude with an outlook to related questions of our findings in section \ref{sec:conclusion_and_outlook}.

\section{\label{sec:energy_eigenstate_dynamics}Relativistic quantum description}

\subsection{Laser setup\label{sec:laser_set_up}}

The setup of consideration in this article is the 2-photon Kapitza-Dirac effect, in which an electron is diffracted in a standing wave of light, given by the vector potential \cite{ahrens_bauke_2012_spin-kde,ahrens_bauke_2013_relativistic_KDE}
\begin{equation}
	\vec{A}(\vec{x},t)=\vec{A}_0\cos(\vec k \cdot \vec x )\sin(\omega t)\,\xi(t)\,.
	\label{eq:vector_potential}
\end{equation}
Here, we introduce the laser wave vector $\vec k = (k_L,0,0)$, the laser frequency $\omega = c k_L$ and a temporal envelope function
\begin{align}
	\xi (t)=
\begin{cases}
	\sin^2(\frac{\pi t}{2 \Delta T}) &\text{ if $0\le t\le \Delta T$,}\\
	1	&\text{ if $\Delta T\le t\le T- \Delta T$,}\\
	\sin^2(\frac{\pi (T-t)}{2 \Delta T}) &\text{ if $T- \Delta T\le t\le  T$,}\\
	0&\text{ else,}
\end{cases}
	\label{eq:temporal_envelope_function}
\end{align}
for modeling the turn on and turn off of the standing light wave. In the envelope function, we also denote the turn on and turn off time $\Delta T$ and the total interaction time $T$. We further mention, that we are working in a Gaussian unit system for all calculations in this article, with the vacuum speed of light $c$, the electron mass $m$, the elementary charge $e$ and the reduced Planck constant $\hbar$. The polarization and amplitude of the standing light wave is parameterized by the vector $\vec A_0=(0,0,A_0)$, which is pointing along the $\vec e_3$ direction.

\subsection{Dirac equation}\label{sec:dirac_equation}

We start our investigation of the coupling to negative states with introducing the relativistic quantum description for spin 1/2 particles, which is formulated in terms of the Dirac equation. We will later draw an analogy to non-relativistic quantum dynamics in terms of the Schr\"{o}dinger equation and after that further consider the relation to classical mechanics.

We denote the Dirac equation by
\begin{equation}
	i\hbar\dot{\Psi}(\vec{x},t)= H\Psi(\vec{x},t)
	\label{eq:dirac_equation}
\end{equation}
with the Dirac Hamiltonian \cite{wachter_2011_relativistic_quantum_mechanics_2011,schwabl_2000_advanced_quantum_mechanics}
\begin{equation}
	H = c\vec \alpha\left(\vec p -\frac{e}{c}\vec A\right) + \beta mc^2\,,
	\label{eq:hamiltonian_dirac_s}
\end{equation}
where we assume the non-zero vector potential in Eq. \eqref{eq:vector_potential}, in this article. The $4\times4$ matrices $\vec{\alpha} = (\alpha_1,\alpha_2,\alpha_3)$\textsuperscript{T} and $\beta$ are
\begin{equation}
	\vec \alpha_i=
	\begin{pmatrix}
		0&\sigma_i\\
		\sigma_i&0
	\end{pmatrix},\quad
	\beta = 
	\begin{pmatrix}
		\mathds{1}&0\\
		0&-\mathds{1}
	\end{pmatrix}\,,
\end{equation}
with the $2\times2$ identity $\mathds{1}$ and the Pauli matrices $\sigma_i$. %
The Dirac Hamiltonian \eqref{eq:hamiltonian_dirac_s} can be decomposed into a free Hamiltonian $H_0$ and an interaction part $V$
\begin{subequations}%
	\begin{align}%
		H_0 &= c\vec\alpha\cdot \vec p +\beta mc^2\label{eq:free_hamiltonian_s}\\
		V&=-e\vec \alpha \cdot\vec A\label{eq:interaction_i}\,.
	\end{align}%
\end{subequations}%
Furthermore, in analogy to previous investigations \cite{ahrens_bauke_2012_spin-kde,ahrens_bauke_2013_relativistic_KDE,ahrens_2017_spin_filter,ahrens_2020_two_photon_bragg_scattering}, the Dirac wave function $\Psi(\vec x ,t)$ is expanded into plane waves
\begin{equation}
	\Psi(\vec x ,t) =\sum_{n,\gamma,s}c^{\gamma,s}_{n}(t)\psi^{\gamma,s}_n(\vec x )
	\label{eq:wave_function_expand}
\end{equation}
with basis elements
\begin{equation}
	\psi^{\gamma,s}_n(\vec x )=\sqrt{\frac{k_L}{2\pi}}u^{\gamma,s}_ne^{i\vec p_n \cdot \vec x/\hbar}\,,\label{eq:bi_spinor_plane_waves}
\end{equation}
electron polarizations $\gamma \in \{+,-\}$, spin polarizations $s \in \{\uparrow,\downarrow\}$ and discrete momenta $\vec{p}_n=\vec{p}_0+n\hbar\vec{k}_L$. The number $n$ represents the number of laser photon momenta along the laser propagation direction in momentum space and the complex amplitudes $c^{\gamma,s}_{n}(t)$ serve as expansion coefficients for $\Psi(\vec x ,t)$ with respect the basis functions $\psi^{\gamma,s}_n(\vec x)$. The explicit form of the bi-spinors $u^{\gamma,s}_{n}=u^{\gamma,s}(\vec p_n)$ is given by
\begin{subequations}
  \begin{equation}
		u^{+,s}(\vec p_n)=\sqrt{\frac{\mathcal{E}_n+mc^2}{2\mathcal{E}_n}}
		\begin{pmatrix}
			\chi^{s} \\
			\frac{\vec{\sigma}\cdot c\vec{p}_n}{\mathcal{E}_n+mc^2}\chi^{s}
		\end{pmatrix}
	\end{equation}
	and
	\begin{equation}
		u^{-,s}(\vec p_n)=\sqrt{\frac{\mathcal{E}_n+mc^2}{2\mathcal{E}_n}}
		\begin{pmatrix}
			-\frac{\vec{\sigma}\cdot c\vec{p}_n}{\mathcal{E}_n+mc^2}\chi^{s}\\
			\chi^{s}
		\end{pmatrix}
	\end{equation}
	\label{eq:bispinors_form}
\end{subequations}
with the two spin polarizations
\begin{equation}
	\chi^{\uparrow}=
	\begin{pmatrix}
		1\\0
	\end{pmatrix}\quad\quad\quad
	\chi^{\downarrow}=
	\begin{pmatrix}
		0\\1
	\end{pmatrix}\,.
\end{equation}
We also point out that the basis functions $\psi^{\gamma,s}_n(\vec x)$ are eigenfunctions of the free, time-independent Dirac Hamiltonian $H_0$ with energy eigenvalue
\begin{equation}
	\mathcal{E}_n = \sqrt{m^2c^4+\vec{p}_n^2c^2}\label{eq:relativistic_energy_momentum_relation}
\end{equation}
and frequencies $\omega_n = \mathcal{E}_n / \hbar$\,.

\subsection{Numerical solution}\label{sec:numerical_solution_dirac_equation}
An equation of motion for the expansion coefficients $c^{\gamma,s}_{n}(t)$ can be obtained by inserting the wave function expansion \eqref{eq:wave_function_expand} into the Dirac equation \eqref{eq:dirac_equation} and projecting with the basis functions \eqref{eq:bi_spinor_plane_waves} from the left. The projection operation can be implemented by integration over one wave length $\lambda=2 \pi/k_L$ along the laser direction, resulting in the momentum space notion for the Dirac equation \cite{ahrens_2012_phdthesis_KDE,ahrens_bauke_2012_spin-kde,ahrens_bauke_2013_relativistic_KDE}
\begin{align}
	i\hbar \dot c^{\gamma,s}_{n}(t)
	=& \pm \mathcal{E}_nc^{\gamma,s}_{n}(t) + \sum_{n^\prime,\gamma^\prime,s^\prime} V^{\gamma,s;\gamma^\prime,s^\prime}_{n,n^\prime}(t) c^{\gamma^\prime,s^\prime}_{n^\prime}(t)\,.
	\label{eq:inserted_dirac_equation}
\end{align}
At this point we have introduced the interaction matrix elements in the Schr\"{o}dinger picture
\begin{align}
	&V_{n,n^\prime}^{\gamma,s;\gamma^{\prime},s^\prime}(t)\nonumber\\
	=&\frac{1}{\lambda}\int_{-\lambda/2}^{\lambda/2} dx \, \left[\psi^{\gamma,s}_{n}(\vec x )\right]^\dagger[-e\vec \alpha\cdot \vec A(\vec x,t)]\psi^{\gamma^\prime,s^\prime}_{n^\prime}(\vec x )\nonumber\\
	=&-\frac{e A_0 \sin(\omega t)\xi(t)}{2} L_{n,n'}^{\gamma,s;\gamma',s'} \delta_{n,n^\prime+1}\nonumber\\
	 &-\frac{e A_0 \sin(\omega t)\xi(t)}{2} L_{n,n'}^{\gamma,s;\gamma',s'} \delta_{n,n^\prime-1}
	\label{eq:interaction_matrix}
\end{align}
and the spinor-matrix elements
\begin{equation}%
 L_{n,n'}^{\gamma,s;\gamma',s'} = \left[(u^{\gamma,s}_n)^\dagger \alpha_3 (u^{\gamma^\prime,s^\prime}_{n^\prime})\right]\,.
 \label{eq:spinor_matrix}
\end{equation}%

\begin{figure}
	\includegraphics[width=0.48 \textwidth]{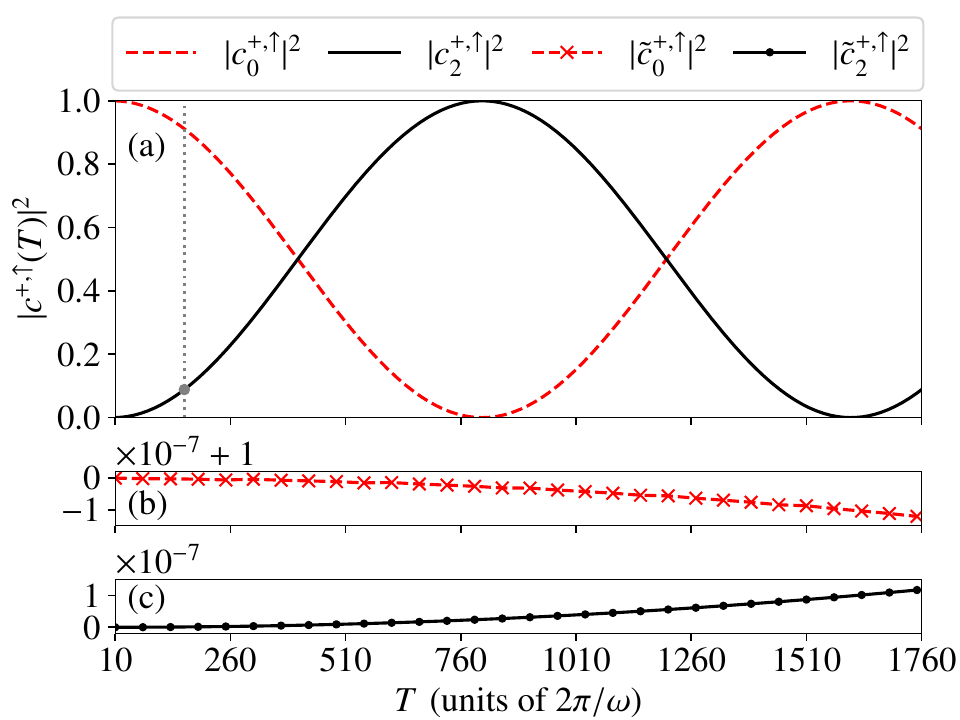}
	\caption{\label{fig:time_evolution}Occupation probabilities as a function of the full interaction time $T$ for the dynamics of the simulated the two-photon Kapitza-Dirac effect. (a) Displayed are the absolute value squares of the only non-vanishing expansion coefficients $c^{+\uparrow}$ and $c^{-\uparrow}$ for a simulation of the equation of motion \eqref{eq:inserted_dirac_equation} with initial condition \eqref{eq:FSME_initial_conditions} and parameters $k_L=0.02\,mc/\hbar$, $\Delta T=5 \times 2 \pi/\omega$ and $eA_0 = 0.01\,mc^2$. The observed Rabi period of 1600 laser cycles can be  deduced analytically, as done in Eq. \eqref{eq:rabi_period} by a calculation using time dependent perturbation theory. The gray dotted line marks the simulation time at which the diffraction probability as a function of the transverse momentum $p_3$ is displayed in Fig. \ref{fig:Simulation_and_Predicted}. Panels (b) and (c) show the quantities $|\tilde c_0^{+,\uparrow}|^2$ and $|\tilde c_2^{+,\uparrow}|^2$, respectively, for which the simulation in panel (a) is redone with the matrix entries $V^{\pm,s;\mp,s^\prime}_{n,n^\prime}$ set to zero.}
\end{figure}

The equation of motion in momentum space \eqref{eq:inserted_dirac_equation} allows for studying the influence of negative state coupling on the quantum dynamics of the electron, by toggling the interactions $V^{\pm,s;\mp,s^\prime}_{n,n^\prime}$ on and off.
We illustrate this for the case of a numerical solution of Eq. \eqref{eq:inserted_dirac_equation}, which is displayed in Fig. \ref{fig:time_evolution}(a), where we set the laser frequency $\omega=c k_L=0.02\,mc^2/\hbar$, the electron initial momentum $\vec p_0=(- \hbar k_L,0,0)$ and the initial quantum state
\begin{equation}
	c_0^{+,\uparrow}(0) = 1\,,\qquad c_n^{\gamma,s}(0)=0\,, \textrm{ else.}\label{eq:FSME_initial_conditions}
\end{equation}
The quantum state is evolved in the external standing wave laser field \eqref{eq:vector_potential} parameterized by a turn-on and turn-off time of ten laser periods $\Delta T=5 \times 2 \pi/\omega$, with the laser field amplitude $eA_0 = 0.01\,mc^2$. We observe in Fig. \ref{fig:time_evolution}(a) sinuously oscillating occupation probabilities, which are consistent with the literature \cite{batelaan_2000_KDE_first,batelaan_2007_RMP_KDE}, and which are termed Rabi oscillations \cite{Scully_Zubairy_1997}. In appendix \ref{sec:appendix_weak_perturbative} we further investigate the weakly non-perturbative regime for clarifying the validity of the perturbative treatment in the sections below, where the perturbation treatment is with respect to the field amplitude parameter $e A_0/m c^2$.

When redoing the simulation with the matrix entries $V^{\pm,s;\pm,s^\prime}_{n,n^\prime}$ set to zero, we obtain dynamics which appear to be identical to the Rabi oscillations as shown in Fig. \ref{fig:time_evolution}(a). Oppositely, we observe a drastic change of the dynamics when setting the matrix entries $V^{\pm,s;\mp,s^\prime}_{n,n^\prime}$ to zero in Eq. \eqref{eq:inserted_dirac_equation}, as shown in Figs. \ref{fig:time_evolution}(b) and \ref{fig:time_evolution}(c). We see in Figs. \ref{fig:time_evolution}(b) and \ref{fig:time_evolution}(c) the beginning of Rabi cycles with a Rabi period which is orders of magnitudes larger than the Rabi oscillation period in Fig. \ref{fig:time_evolution}(a). Therefore, we are led to the conclusion that the couplings between negative and positive states $V^{\pm,s;\mp,s^\prime}_{n,n^\prime}$ contribute significantly more to the quantum dynamics than the couplings $V^{\pm,s;\pm,s^\prime}_{n,n^\prime}$, which only mediate within the positive and negative spectrum.

\begin{figure}
	\includegraphics[width=0.48 \textwidth]{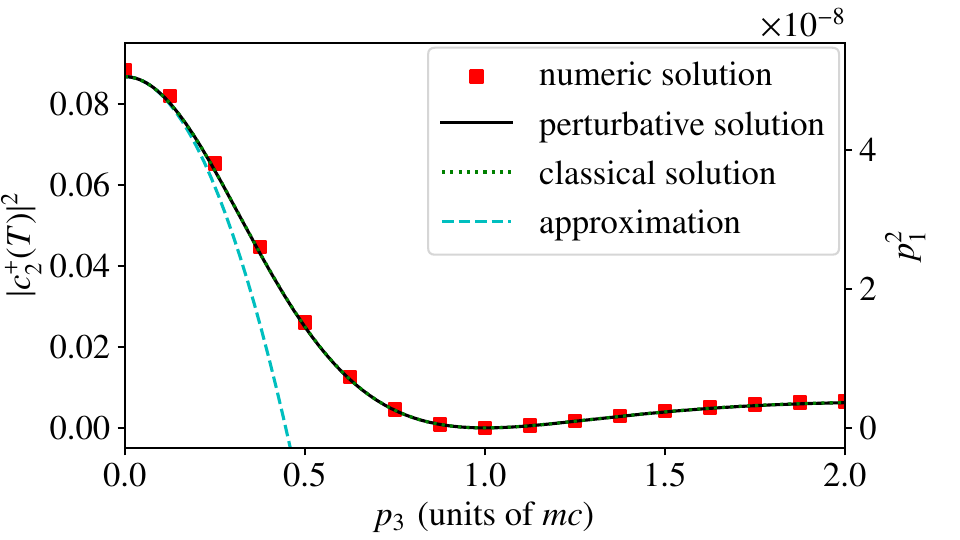}
	\caption{\label{fig:Simulation_and_Predicted}Diffraction probability \eqref{eq:c_plus_diffraction_probablity} of the two-photon Kapitza-Dirac effect as a function of the transverse electron momentum $p_3$. The parameters and initial condition \eqref{eq:FSME_initial_conditions} are the same as for Fig. \ref{fig:time_evolution} with the specific interaction time $T=150 \times 2\pi/\omega$, where the first data point of the numeric solution at $p_3=0$ corresponds to the gray dot at the gray dotted time mark in Fig. \ref{fig:time_evolution}(a). The results of the numeric solutions of the differential equation \eqref{eq:inserted_dirac_equation} are marked by red boxes and the analytic short-time solution from time-dependent perturbation theory (perturbative solution) of Eq. \eqref{eq:coherent_sum_expression} is displayed by a solid black line. The green dotted line marks the square of the electron's classical momentum in laser propagation direction (axis on the right) for the case of the relativistic description of the classical equations of motion \eqref{eq:relativistic_motion_equation}. This classical solution is performed with a vector potential amplitude of $eA_0 = 10^{-3}\,mc^2$, with all other parameters being consistent with the values used for the quantum calculations. The cyan dashed line marks the low momentum approximation \eqref{eq:dirac_perturbative_diffraction_probability_approximation} of the perturbative relativistic quantum solution, where we mention that the relativistic classical solution \eqref{eq:relativistic_ponderomotive_force} exhibits the same scaling with respect to the transverse momentum $p_3$ as the analytic quantum solution.}
\end{figure}

In order to further identify characteristic features of the diffraction probability in a later comparison with perturbative results, we display the momentum dependence of the resulting diffraction probability
\begin{equation}
 |c_2^+(t)|^2 = |c_2^{+,\uparrow}(t)|^2 + |c_2^{+,\downarrow}(t)|^2 \label{eq:c_plus_diffraction_probablity}
\end{equation}
for multiple simulations with different transverse momenta $p_3$ for the initial momentum $\vec p_0=(- \hbar k_L, 0, p_3)$ at time $T=150 \times 2\pi/\omega$ in Fig. \ref{fig:Simulation_and_Predicted}.

\subsection{Perturbative solution}\label{sec:perturbative_solution_two_kde_dirac}

For a more detailed study of the contribution of negative energy eigenstates in the quantum dynamics as shown in Figs. \ref{fig:time_evolution} and \ref{fig:Simulation_and_Predicted}, we are now discussing the perturbative short-time solution for the description of the process here, by following the conventions in references \cite{ahrens_2012_phdthesis_KDE,ahrens_bauke_2013_relativistic_KDE,ahrens_2020_two_photon_bragg_scattering,ahrens_guan_2022_beam_focus_longitudinal}. The diffraction amplitude of the two-photon Kapitza-Dirac effect can be encoded by the time-evolution of the two-component object
\begin{equation}\label{eq:relativistic_spinor_state}
 c^+_n(t)=
\begin{bmatrix}
 c_n^{+,\uparrow}(t)\\
 c_n^{+,\downarrow}(t)
\end{bmatrix}\,,
\end{equation}
where its absolute value square is already denoted in Eq. \eqref{eq:c_plus_diffraction_probablity}. We again assume the initial condition \eqref{eq:FSME_initial_conditions} for our system. Then $c^+_2(t)$ is related to the initial state $c^{+,\uparrow}_0(t)$ by the sub-matrix element $U^{+s;+\uparrow}_{2,0}(t,0)$ of the time-evolution operator via the general propagation equation
\begin{equation}
	c^{\gamma,s}_n(t) = \sum_{n',\gamma',s'} U^{\gamma,s;\gamma',s'}_{n,n'}(t,0)\,c^{\gamma',s'}_{n'}(0) \,.\label{eq:pertubation_theory}
\end{equation}
The next neighbor-coupling of the interaction \eqref{eq:interaction_matrix} implies the necessity for second order time-dependent perturbation theory for the lowest non-vanishing contribution in $U^{+s;+\uparrow}_{2,0}(t,0)$. The perturbative time-evolution operator can be denoted as \cite{sakurai2014modern}
\begin{equation}
	U(t,0)=\left(-\frac{i}{\hbar}\right)^2\int^t_0 \,dt^\prime\int^{t^\prime}_0 \,dt^{\prime\prime} \tilde V(t^{\prime}) \tilde V(t^{\prime\prime})\,,
	\label{eq:general_second_perturbation_theory}
\end{equation}
with the interaction $V$ of Eq. \eqref{eq:interaction_matrix}, in the interaction picture
\begin{align}
	\tilde V = e^{iH_0t/\hbar}V e^{-iH_0t/\hbar}\,.\label{eq:transfer_interaction_shcrodinger_picture}
\end{align}
In explicit index notion this interaction term reads as
\begin{align}
	\tilde V_{n,n^\prime}^{\gamma,s;\gamma^\prime,s^\prime}=V_{n,n^\prime}^{\gamma,s;\gamma^\prime,s^\prime} e^{i(\gamma\omega_n-\gamma^\prime\omega_{n^\prime})t}\,,
	\label{eq:transfer_interaction_shcrodinger_picture_matrix_element}
\end{align}
and allows us to denote Eq. \eqref{eq:general_second_perturbation_theory} in the form
\begin{subequations}\label{eq:propagator_process}%
\begin{align}%
	U^{+,s;+,\uparrow}_{2,0}(t,0)
	&= \sum_{\gamma^\prime,s^\prime}\left(-\frac{i}{\hbar}\right)^2 \int^t_0 \,dt^\prime \int^{t^\prime}_0\,dt^{\prime\prime}\\
	&\quad \times e^{i(\omega_2-\gamma^\prime\omega_{1})t^\prime}V_{2,1}^{+,s;\gamma^\prime,s^\prime}(t^\prime)\\
	&\quad \times e^{i(\gamma'\omega_1-\omega_{0})t^{\prime\prime}}V_{1,0}^{\gamma^\prime,s^\prime;+,\uparrow}(t^{\prime\prime})\,.
\end{align}
\end{subequations}
We assume $\xi(t) = 1$ for the situation of an assumed short turn on and turn off of the standing light wave ($\Delta T \ll T$) in Eq. \eqref{eq:temporal_envelope_function}. By substituting Eq. \eqref{eq:interaction_matrix} into Eq. \eqref{eq:propagator_process} we obtain
\begin{subequations}\label{eq:propagator_without_integration}%
\begin{align}%
	U^{+,s;+,\uparrow}_{2,0}(t,0) &= \left(-\frac{i}{\hbar}\frac{e A_0}{4}\right)^2 \\
	& \qquad\times \sum_{\gamma^\prime,s^\prime} \Xi^{\gamma'} L_{2,1}^{+,s;\gamma',s'} L_{1,0}^{\gamma',s';+,\uparrow}\,,
\end{align}%
\end{subequations}%
with the double time-integral
\begin{multline}
 \Xi^{\gamma} = - \int^t_0 \,dt^\prime \int^{t^\prime}_0\,dt^{\prime\prime} e^{i(\omega_2-\gamma \omega_{1})t^\prime} \left(e^{i\omega t^{\prime}}-e^{-i\omega t^{\prime}}\right) \\
 \times e^{i(\gamma \omega_1-\omega_{0})t^{\prime\prime}} \left(e^{i\omega t^{\prime\prime}}-e^{-i\omega t^{\prime\prime}}\right)\,.\label{eq:dirac_perturbation_time_integral}
\end{multline}
Relevant for the population transfer of Rabi cycles in the Kapitza-Dirac effect are the non-oscillatory terms, which is a reason for us, to neglect the counter-rotating terms $e^{\pm i\omega t^{\prime}}e^{\pm i\omega t^{\prime\prime}}$ and the lower limit of the $t''$ integral \cite{ahrens_bauke_2013_relativistic_KDE,ahrens_2020_two_photon_bragg_scattering,ahrens_guan_2022_beam_focus_longitudinal} in Eq. \eqref{eq:dirac_perturbation_time_integral}. Then, for each $\gamma\in\{+,-\}$ the integrand in \eqref{eq:dirac_perturbation_time_integral} contains the product of phases
\begin{equation}
 e^{i(\omega_2-\gamma \omega_{1}-a \omega)t^\prime} e^{i(\gamma \omega_1-\omega_{0}+a \omega)t^{\prime\prime}}\,,\label{eq:KDE_time-integral_exponentials}
\end{equation}
with the two possibilities $a\in\{-,+\}$. The upper limit of the $t''$ integral evaluates to $- i F^\gamma$, with
\begin{subequations}\label{eq:two_coefficients_l}
\begin{equation}
F_a^\gamma =\frac{1}{\gamma\omega_1 - \omega_0+ a\omega}\,,\label{eq:two_coefficients_l_photons}
\end{equation}
and
\begin{equation}
F^\gamma = F_{+}^\gamma + F_{-}^\gamma \,.\label{eq:two_coefficients_l_sum}
\end{equation}
\end{subequations}
Note, that $\omega_2=\omega_0$ holds and the exponential in Eq. \eqref{eq:KDE_time-integral_exponentials} cancels, when integrating \eqref{eq:dirac_perturbation_time_integral}, due to the momentum configuration which we have imposed for obtaining classical energy and momentum conservation of the final states in the Kapitza-Dirac effect \cite{ahrens_bauke_2012_spin-kde,ahrens_bauke_2013_relativistic_KDE}. The second $t'$ integral then yields $\Xi^{\gamma}=-i F^\gamma t$ in Eq. \eqref{eq:dirac_perturbation_time_integral}, such that Eq. \eqref{eq:propagator_without_integration} can be written as
\begin{subequations}\label{eq:propagator_after_integration}%
\begin{align}%
	U^{+,s;+,\uparrow}_{2,0}(t,0) &= i \left(\frac{e A_0}{4 \hbar}\right)^2 t \\
	&\qquad \times \sum_{\gamma^\prime,s^\prime} F^{\gamma'} L_{2,1}^{+,s;\gamma',s'} L_{1,0}^{\gamma',s';+,\uparrow}\,.\label{eq:propagator_after_integration_coupling_part}
\end{align}%
\end{subequations}%

\begin{figure}%
	\includegraphics[width=0.42 \textwidth]{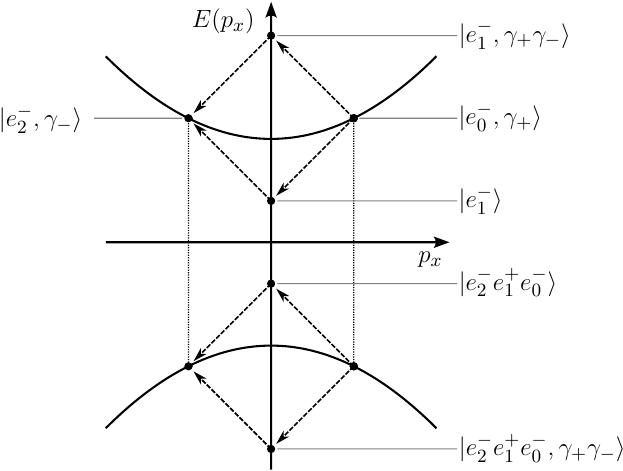}
	\caption{\label{combined_dispersion} Coupling paths in perturbation theory which are corresponding to the four intermediate states in Eq. \eqref{eq:intermediate_quantized_states} in the context of Compton scattering \cite{ahrens_2020_two_photon_bragg_scattering}. The interaction starts with the single electron $e_0^-$, single photon $\gamma_+$ state $\ket{e^-_0,\gamma_+}$ and also ends with a single electron $e_2^-$, single photon $\gamma_-$ state $\ket{e^-_2,\gamma_-}$, where the electron acquires two photon momenta $0\rightarrow 2$ and the photon momentum is reversed $+\rightarrow -$. The quantum paths in the negative energy continuum (corresponding to $F_a^-$) are associated with the state configuration of an electron plus an electron-positron pair $\ket{e^-_2 e^+_1 e^-_0}$, $\ket{e^-_2 e^+_1 e^-_0,\gamma_+\gamma_-}$ in a fully quantized theory.}
\end{figure}

This result can be associated with a semi-classical interaction picture \cite{ahrens_bauke_2012_spin-kde,ahrens_bauke_2013_relativistic_KDE} as illustrated in Fig. \ref{combined_dispersion}, where the initial quantum state $c_0^+$ and final state $c_2^+$ correspond to
\begin{subequations}\label{eq:initial_and_final_quantized_states}%
\begin{align}%
 \ket{e^-_0,\gamma_+} &= c^\dagger_{\vec p_0} a^\dagger_k \ket{0}\quad\textrm{(initial)}\\
 \ket{e^-_2,\gamma_-} &= c^\dagger_{\vec p_2} a^\dagger_{-k} \ket{0}\quad\textrm{(final)}
\end{align}%
\end{subequations}%
in the context of a fully quantized theory \cite{ahrens_2020_two_photon_bragg_scattering} of the equivalent scattering amplitudes in Compton scattering, in the context of old-fashioned perturbation theory \cite{Halzen_Martin_1984_quarks_and_leptons,Weinberg_1995_quantum_theory_of_fields}. The right-hand side of Eq. \eqref{eq:initial_and_final_quantized_states} denotes the vacuum state $\ket{0}$, photon creation operator $a^\dagger_k$ and electron creation operator $c^\dagger_{\vec p_n}$ at momentum $n$, where we have omitted spin- and polarization degrees of freedom, for simplicity. Carrying on as in reference \cite{ahrens_2020_two_photon_bragg_scattering}, the intermediate states in Eq. \eqref{eq:propagator_after_integration} with prefactors $F_a^\gamma$ correspond to the quantized states
\begin{subequations}\label{eq:intermediate_quantized_states}%
\begin{alignat}{2}%
 F_-^+ & \hspace{0.3 cm} \Leftrightarrow \hspace{0.3 cm} \ket{e^-_1} &&= c^\dagger_{\vec p_1} \ket{0} \\
 F_+^+ & \hspace{0.3 cm} \Leftrightarrow \hspace{0.3 cm} \ket{e^-_1 ,\gamma_+\gamma_-} &&= c^\dagger_{\vec p_1} a^\dagger_k a^\dagger_{-k} \ket{0} \\
 F_+^- & \hspace{0.3 cm} \Leftrightarrow \hspace{0.3 cm} \ket{e^-_2 e^+_1 e^-_0} &&= c^\dagger_{\vec p_2} d^\dagger_{\vec p_1} c^\dagger_{\vec p_0} \ket{0} \\
 F_-^- & \hspace{0.3 cm} \Leftrightarrow \hspace{0.3 cm} \ket{e^-_2 e^+_1 e^-_0,\gamma_+\gamma_-} &&= c^\dagger_{\vec p_2} d^\dagger_{\vec p_1} c^\dagger_{\vec p_0} a^\dagger_k a^\dagger_{-k} \ket{0}
\end{alignat}%
\end{subequations}%
with positron creation operators $d^\dagger_{\vec p_n}$. In this context we mention that electronic anti-particle states are added with their positive energy to a quantized Hamiltonian in quantum field theory, by a canonical quantization procedure, for obtaining a Hamiltonian in the theory which is bound from below. Therefore, the excitation energies of the photon excitations are appearing mirrored along the energy axis in the classical single particle energy-momentum diagram in Fig. \ref{combined_dispersion}.

\subsection{Coupling to negative states\label{sec:negative_state_investigation}}

Having discussed the physical interpretation of the perturbative solution \eqref{eq:propagator_after_integration}, we now draw our attention to the aspired coupling strength of the quantum paths \eqref{eq:intermediate_quantized_states}, which contribute to the diffraction probability. The diffraction probability $|c_2^+(T)|^2$ in Eq. \eqref{eq:c_plus_diffraction_probablity} can be expanded by using the propagation equation \eqref{eq:pertubation_theory} with inserting the matrix element from the lowest perturbative contribution \eqref{eq:propagator_after_integration} resulting in
\begin{subequations}\label{eq:coherent_sum_expression}%
\begin{align}%
 &\left|c_2^+(T)\right|^2 = \sum_{s}\left|U^{+,s;+,\uparrow}_{2,0}(T,0)\right|^2\label{eq:diffraction_probability_propagator_relation}\\
 &= \sum_{s}\left|\sum_{\gamma^\prime,s^\prime} \left(\frac{e A_0}{4 \hbar}\right)^2 T\,
  F^{\gamma'} L_{2,1}^{+,s;\gamma',s'} L_{1,0}^{\gamma',s';+,\uparrow}\right|^2\,.
\end{align}%
\end{subequations}%
We display this diffraction probability as black solid line in Fig. \ref{fig:Simulation_and_Predicted} as a function of the transverse momentum $p_3$ and find a match with the corresponding numeric solution (red boxes). The observed agreement between numeric and analytic plane-wave solutions of the relativistically described Kapitza-Dirac effect is in line with previous studies \cite{ahrens_2012_phdthesis_KDE,ahrens_bauke_2012_spin-kde,ahrens_bauke_2013_relativistic_KDE,ahrens_2020_two_photon_bragg_scattering}.

\begin{figure}
	\includegraphics[width=0.48 \textwidth]{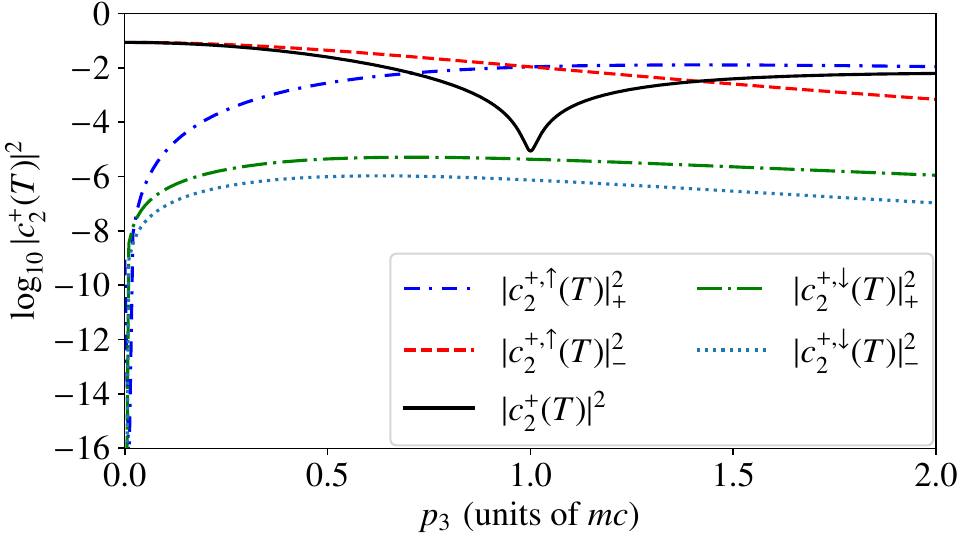}
	\caption{\label{fig:constituents_two_photon} Diffraction probabilities for different coupling paths of the two-photon Kapitza-Dirac effect as a function of the transverse momentum $p_3$. Displayed is the perturbatively solved diffraction probability $|c_2^+(T)|^2$ from Eq. \eqref{eq:coherent_sum_expression} and the spin resolved positive and negative coupling strengths $|c_2^{+,s}(T)|^2_{\gamma'}$ from Eq. \eqref{eq:individual_coupling}. The probability $|c_2^+(T)|^2$ is the same as the black line in Fig. \ref{fig:Simulation_and_Predicted}, where its shape appears differently due to the logarithmic scaling in this figure. We observe that the coupling path with a negative intermediate electron state $|c_2^{+,\uparrow}(T)|^2_{-}$ is dominating the diffraction probability for $p_3<mc$\,.}
\end{figure}

The advantage of the perturbative solution is that it allows for isolating the contribution which involves the coupling to virtual negative states in the dynamics. To do so, we now remove the sum over the intermediate electron energy given by the index $\gamma'$ and also the spin sum over the index $s$ in the total diffraction probability in Eq. \eqref{eq:coherent_sum_expression} and define this as the quantity
\begin{equation}\label{eq:individual_coupling}
 \left|c_2^{+,s}(T)\right|^2_{\gamma'} = \left|\sum_{s^\prime} \left(\frac{e A_0}{4 \hbar}\right)^2 T\,
  F^{\gamma'} L_{2,1}^{+,s;\gamma',s'} L_{1,0}^{\gamma',s';+,\uparrow}\right|^2\,.
\end{equation}
The object $|c_2^{+,s}(T)|^2_{\gamma'}$ with the added $\gamma'$ index provides a measure about how much the positive and negative coupling path contributes to the diffraction probability, when they were not coherently interfering at detection time $T$ of the process. The diffraction probability $|c_2^+(T)|^2$ and the coupling strengths $|c_2^{+,s}(T)|^2_{\gamma'}$ are depicted as a function of the transverse momentum $p_3$ in Fig. \ref{fig:constituents_two_photon}. We see that the coupling $|c_2^{+,\uparrow}(T)|^2_{-}$ of the intermediate negative electron states can dominate over the coupling $|c_2^{+,\uparrow}(T)|^2_{+}$ of the intermediate positive electron states by orders of magnitude, for transverse electron momenta $|p_3|$ along the laser polarization direction smaller than $m c$. At this point we also would like to mention that the two spin-preserving transitions (from initial spin $s'=\uparrow$ to final spin $s=\uparrow$, red dashed and blue dotted lines of the functions $|c_2^{+,\uparrow}(T)|^2_{\pm}$) cancel each other at $p_3=m c$. This cancellation results in a dip in the diffraction probability  $|c_2^+(T)|^2$, where only the spin-changing couplings $|c_2^{+,\downarrow}(T)|^2_{+}$ and $|c_2^{+,\downarrow}(T)|^2_{-}$ contribute to the quantum dynamics, which is the reason as to why a spin-flip in the Kapitza-Dirac effect \cite{ahrens_bauke_2013_relativistic_KDE} can emerge as an observable feature.

We further would like to point out once more, that the dominant coupling $|c_2^{+,\uparrow}(T)|^2_{-}$ to the negative states at small transverse momenta $p_3 < mc$ corresponds to the processes with a virtual electron-positron pair
\begin{subequations}\label{eq:negative_state_coupling_fully_quantized}%
\begin{alignat}{2}%
  \ket{e^-_0,\gamma_+} &\rightarrow \ket{e^-_2 e^+_1 e^-_0}                  &&\rightarrow \ket{e^-_2,\gamma_-}\\
  \ket{e^-_0,\gamma_+} &\rightarrow \ket{e^-_2 e^+_1 e^-_0,\gamma_+\gamma_-} &&\rightarrow \ket{e^-_2,\gamma_-}\,,
\end{alignat}%
\end{subequations}%
for the case of an ultra-low field interaction with a single photon in Compton scattering for the case of a quantized theory \cite{ahrens_2020_two_photon_bragg_scattering} [see particle quantum excitations introduced in Eqs. \eqref{eq:initial_and_final_quantized_states} and \eqref{eq:intermediate_quantized_states}]. The electron-positron pair in Eq. \eqref{eq:negative_state_coupling_fully_quantized} is missing in the processes of intermediate \emph{positive} state coupling amplitude $|c_2^{+,\uparrow}(T)|^2_{+}$, which corresponds to
\begin{subequations}\label{eq:positive_state_coupling_fully_quantized}%
\begin{alignat}{2}%
  \ket{e^-_0,\gamma_+} &\rightarrow \ket{e^-_1}                  &&\rightarrow \ket{e^-_2,\gamma_-}\\
  \ket{e^-_0,\gamma_+} &\rightarrow \ket{e^-_1 ,\gamma_+\gamma_-} &&\rightarrow \ket{e^-_2,\gamma_-}\,.
\end{alignat}%
\end{subequations}%

\subsection{Low transverse momentum approximation}\label{sec:low_transverse_momentum_approximation}

We now establish explicit expressions for the diffraction probability for the case of low transverse electron momenta, for comparison with the non-relativistic and classical descriptions of the quantum dynamics of the 2-photon Kapitza-Dirac effect. In order to do so, we introduce the quantities
\begin{subequations}%
\begin{align}\label{eq:g_pm_coefficients}%
	g^+_n&:=\frac{1}{\sqrt{2}}\left(1+\frac{mc^2}{\mathcal{E}_n}\right)^{1/2}\\
	g^-_n&:=\frac{c}{\sqrt{2}}\left(\frac{1}{\mathcal{E}_n(\mathcal{E}_n+mc^2)}\right)^{1/2}\,,
\end{align}%
\end{subequations}%
and based on them also
\begin{subequations}\label{eq:functions_tsrw}
\begin{align}%
	t_{n,n^\prime} &= g^+_n g^+_{n'}+\vec p_n \cdot\vec p_{n^\prime} \,g^-_n g^-_{n'}\label{eq:function_t}\\
	s^l_{n,n^\prime} &= p_{nl}\,g^-_n g^+_{n'} + p_{n^{\prime}l}\,g^+_n g^-_{n'}\\
	r^l_{n,n^\prime} &= p_{nl}\,g^-_n g^+_{n'} - p_{n^{\prime}l}\,g^+_n g^-_{n'}\\
	w^{lq}_{n,n^\prime} & = p_{nl}\, p_{n^\prime q}\,g^-_n g^-_{n'} + p_{nq}  \,p_{n^\prime l}\,g^-_n g^-_{n'}\,,
\end{align}%
\end{subequations}%
where we denote the $l$-th component of the momentum $\vec p_n$ as $p_{nl} = \vec p_n \cdot \vec e_l$, with the canonical unit vector $\vec e_l$ in $l$-direction. On the basis of these functions, the spinor-matrix contractions \eqref{eq:spinor_matrix} can be expanded as \cite{ahrens_2012_phdthesis_KDE,ahrens_bauke_2013_relativistic_KDE}
\begin{subequations}\label{eq:coupling_matrices_expression}%
	\begin{align}%
		L_{n,n'}^{+,s;+,s'} &= 
		\begin{pmatrix}
			s^3_{n,n^\prime}&-r^1_{n,n^\prime}\\
			r^1_{n,n^\prime}&s^3_{n,n^\prime}
		\end{pmatrix}\quad\label{eq:coupling_matrix_positive}\\
		L_{n,n'}^{\mp,s;\pm,s'} &= 
		\begin{pmatrix}
			t_{n,n^\prime}-w^{33}_{n,n^\prime} &-w^{31}_{n,n^\prime}\\
			-w^{31}_{n,n^\prime}&-t_{n,n^\prime}+w^{33}_{n,n^\prime}
		\end{pmatrix}\,,\label{eq:coupling_matrix_negative}
	\end{align}%
\end{subequations}%
where the columns and rows of the matrices on the right-hand side are indexed by the spinor indices $s'$ and $s$, respectively. With these matrices, the expression in Eq. \eqref{eq:propagator_after_integration_coupling_part} can be reformulated as
\begin{subequations}\label{eq:propagator_coupling_up}%
\begin{align}%
&F^+s^3_{2,1}s^3_{1,0}-F^+r^1_{2,1}r^1_{1,0} \label{eq:propagator_coupling_positive_up}\\
	&+F^-(t_{2,1}-w^{33}_{2,1})(t_{1,0}-w^{33}_{1,0})+F^-w^{31}_{2,1}w^{31}_{1,0} \label{eq:propagator_coupling_negative_up}
\end{align}%
\end{subequations}%
for the final spin index $s=\uparrow$ and
\begin{subequations}\label{eq:propagator_coupling_down}%
\begin{align}%
&F^+s^3_{2,1}r^1_{1,0}+F^+r^1_{2,1}s^3_{1,0} \label{eq:propagator_coupling_positive_down}\\
	&+F^-(t_{2,1}-w^{33}_{2,1})w^{31}_{1,0}-F^-w^{31}_{2,1}(t_{1,0}-w^{33}_{1,0}) \label{eq:propagator_coupling_negative_down}
\end{align}%
\end{subequations}%
for the final spin index $s=\downarrow$. The upper lines in Eqs. \eqref{eq:propagator_coupling_positive_up} and \eqref{eq:propagator_coupling_positive_down} are contributions with the intermediate state energy index $\gamma'=+$ and the lower lines in Eqs. \eqref{eq:propagator_coupling_negative_up} and \eqref{eq:propagator_coupling_negative_down} are contributions with $\gamma'=-$\,. Our aim is now to deduce simplified expressions for Eqs. \eqref{eq:propagator_coupling_up} and \eqref{eq:propagator_coupling_down}, for small momenta $p_3$, with the following hierarchy of magnitudes
\begin{equation}
 \frac{\hbar k_L}{mc} \ll \frac{p_3}{m c} \ll 1\,, \label{eq:order_of_approximations}
\end{equation}
or in a simpler notion $\hbar k_L \ll p_3 \ll m c$. The laser photon energy $c \hbar k_L$ over the electron rest-mass energy $m c^2$ is on the percent level for hard x-ray lasers and on the order of $10^{-6}$ for optical lasers, such that the assumed order \eqref{eq:order_of_approximations} appears justified. Having in mind that $\hbar k_L \ll p_3$, we start with the second order Taylor expansion of $g_n^+$ with respect to $p_3$
\begin{equation}
 g_n^+ \approx 1 - \frac{1}{8}\frac{p_3^2}{m^2 c^2}\,,
\end{equation}
where we later also need the product of these coefficients, which are approximately
\begin{equation}
 g_n^+ g_{n'}^+ \approx 1 - \frac{1}{4}\frac{p_3^2}{m^2 c^2}\,.\label{eq:gp_gp}
\end{equation}
Regarding the coefficients $g_n^-$ we only keep the zeroth order of the series expansion
\begin{equation}
 g_n^- \approx \frac{1}{2 m c}\,,
\end{equation}
which we justify by the observation that the $g_n^-$ only appear together with components of the electron momenta $\vec p_n=[(n-1) \hbar k_L, 0, p_3]^T$. Specifically we denote
\begin{subequations}%
\begin{align}%
 p_{n1} \,g_n^- &\approx \frac{(n-1) \hbar k_L}{2 m c} \label{eq:g-_p1}\\
 p_{n3} \,g_n^- &\approx \frac{p_3}{2 m c}\,.\label{eq:g-_p3}
\end{align}%
\end{subequations}%

Since we have $\hbar k_L \ll p_3$, Eq. \eqref{eq:g-_p1} can be neglected, when compared with Eq. \eqref{eq:g-_p3}. For the same reason we can approximate the product of momenta $\vec p_n$ and $\vec p_{n'}$ as
\begin{equation}
 \vec p_n \cdot \vec p_{n'}= (n-1) (n'-1) \hbar^2 k_L^2 + p_3^2 \approx p_3^2\,,
\end{equation}
and consequently
\begin{equation}
 \vec p_n \cdot \vec p_{n'} \, g_n^- g_{n'}^-\approx \frac{1}{4}\frac{p^2_3}{m^2 c^2}\,.\label{eq:p_dot_p_gm_gm}
\end{equation}
This allows us to approximate \eqref{eq:function_t} up to order $p_3^2$ as
\begin{subequations}%
\begin{equation}
 t_{n,n'} \approx 1\,,\label{eq:t_approximation}
\end{equation}
where Eqs. \eqref{eq:gp_gp} and \eqref{eq:p_dot_p_gm_gm} were substituted. Similarly, the other relevant expressions in Eqs. \eqref{eq:functions_tsrw} can be written as
\begin{align}%
 s^3_{n,n'} &\approx \frac{p_3}{m c}\label{eq:s3_approximation}\\
 r^1_{n,n'} &\approx \frac{(n-n')\hbar k_L}{2 m c}\\
 w^{31}_{n,n'} &\approx \frac{(n+n'-2)\hbar k_L p_3}{4 m^2 c^2}\\
 w^{33}_{n,n'} &\approx \frac{p_3^2}{2 m^2 c^2}\,.\label{eq:w33_approximation}
\end{align}%
\end{subequations}%
It remains to find approximations for the coefficients $F^\gamma$, because the $F^\gamma$ also appear in the expressions \eqref{eq:propagator_coupling_up} and \eqref{eq:propagator_coupling_down}, which we want to approximate. We first note, that we can simplify the sum \eqref{eq:two_coefficients_l_sum} into
\begin{equation}
 F^\gamma
 = \frac{2 \hbar (\gamma \mathcal{E}_1 - \mathcal{E}_0)}{(\gamma \mathcal{E}_1 - \mathcal{E}_0)^2 - \hbar^2 \omega^2}
 = \frac{2 \hbar (\gamma \mathcal{E}_1 - \mathcal{E}_0)}{2 \mathcal{E}_1 (\mathcal{E}_1 - \gamma \mathcal{E}_0)}
 = \gamma \frac{\hbar}{\mathcal{E}_1}\,,\label{eq:F_simplification}
\end{equation}
where we manipulated a term in the denominator into
\begin{equation}
 (\gamma \mathcal{E}_1 - \mathcal{E}_0)^2
 = \mathcal{E}_1^2 + \mathcal{E}_0^2 - 2 \gamma \mathcal{E}_0 \mathcal{E}_1
 = 2 \mathcal{E}_1^2 + \hbar^2 \omega^2 - 2 \gamma \mathcal{E}_0 \mathcal{E}_1
\end{equation}
and inserted a factor $\gamma^2=1$ in the denominator before the last equality in \eqref{eq:F_simplification}. Finally, a Taylor expansion of Eq. \eqref{eq:F_simplification} yields
\begin{equation}
 F^\gamma = \gamma \frac{\hbar}{m c^2}\left( 1 - \frac{1}{2} \frac{p_3^2}{m^2 c^2}\right)\,.\label{eq:F_approximation}
\end{equation}
With these approximations, we observe that the spin-preserving expression \eqref{eq:propagator_coupling_up} dominates by a factor $p_3/(\hbar k_L)$ over the spin flipping expression \eqref{eq:propagator_coupling_down}, such that we assume that \eqref{eq:propagator_coupling_down} can be neglected. Therefore, we can write the dominant contribution in Eq. \eqref{eq:propagator_after_integration} as
\begin{equation}
U^{+,\uparrow;+,\uparrow}_{2,0}(t,0) =-i \frac{e^2 A_0^2t}{16mc^2\hbar}\left( 1 - \frac{5}{2} \frac{p_3^2}{m^2 c^2}\right)\,,\label{eq:dirac_perturbation_approximation}
\end{equation}
by inserting the largest terms in Eqs. \eqref{eq:t_approximation}, \eqref{eq:s3_approximation}, \eqref{eq:w33_approximation} and \eqref{eq:F_approximation} into Eq.  \eqref{eq:propagator_coupling_up}. Eq. \eqref{eq:dirac_perturbation_approximation} allows us to write the short time diffraction probability \eqref{eq:diffraction_probability_propagator_relation} up to quadratic order in $p_3/(mc)$ as
\begin{equation}
 \left|c_2^+(T)\right|^2 \approx \left(\frac{e^2 A_0^2T}{16mc^2\hbar}\right)^2\left( 1 - 5 \frac{p_3^2}{m^2 c^2}\right) \,,\label{eq:dirac_perturbative_diffraction_probability_approximation}
\end{equation}
which we include as cyan dashed line in Fig. \ref{fig:Simulation_and_Predicted}. From the figure we conclude a good agreement between the perturbative solution \eqref{eq:coherent_sum_expression} and its approximation for small transverse momenta \eqref{eq:dirac_perturbative_diffraction_probability_approximation}. We can further confirm the approximation of the short time solution of predicted Rabi oscillations \cite{sakurai2014modern,Scully_Zubairy_1997} in Kapitza-Dirac scattering \cite{batelaan_2000_KDE_first,batelaan_2007_RMP_KDE,ahrens_bauke_2012_spin-kde,ahrens_bauke_2013_relativistic_KDE}
\begin{subequations}%
	\begin{align}%
		|c^{+}_0(t)|^2 &= \cos^2\left(\frac{\Omega_Rt}{2}\right)\\
		|c^{+}_2(t)|^2 &=\sin^2\left(\frac{\Omega_Rt}{2}\right)\,,
	\end{align}
	\label{eq:ansatz}%
\end{subequations}%
where the Rabi frequency at $p_3=0$ can be identified from Eq. \eqref{eq:dirac_perturbation_approximation} as
\begin{equation}
	\Omega_{R} = \frac{e^2 A_0^2}{8mc^2\hbar}\,.\label{eq:rabi_frequency}
\end{equation}
For the parameter $eA_0 = 0.01\,mc^2$ in Fig. \ref{fig:time_evolution} this results in $\Omega_{R}= 1.25\times 10^{-5} \,mc^2/\hbar$. The corresponding Rabi period
\begin{equation}
 T_R=\frac{2 \pi}{\Omega_{R}}=5.03\times 10^5 \frac{\hbar}{m c^2} = 1600 \frac{2 \pi}{\omega}\label{eq:rabi_period}
\end{equation}
matches the oscillation period in Fig. \ref{fig:time_evolution}(a) and is another confirmation for our approximation of the perturbative calculation.

After having verified the validity of the perturbative solution and its approximation, we want to emphasize again that it is the contribution \eqref{eq:propagator_coupling_negative_up} with scaling
\begin{equation}
 F^-(t_{2,1}-w^{33}_{2,1})(t_{1,0}-w^{33}_{1,0}) \approx 1 - \frac{3}{2} \frac{p_3^2}{m^2 c^2}\,,
\end{equation}
which dominates the quantum dynamics for small and in particular zero transverse electron momenta $p_3=0$ in the two-photon Kapitza-Dirac effect. This is of significant relevance, because the term \eqref{eq:propagator_coupling_negative_up} arises from coupling to \emph{negative} states [ie. $\gamma'=-$ in Eq. \eqref{eq:propagator_after_integration_coupling_part}] and is therefore to be associated with the coupling process \eqref{eq:negative_state_coupling_fully_quantized}, in which a virtual electron-positron pair is involved in the context of Compton scattering in a fully quantized theory.

\section{\label{sec:non_relativistic_quantum_description}Non-relativistic quantum description}

\subsection{\label{sec:schrodinger_equation}Schr\"{o}dinger equation}

In the following, we want to compare corresponding expressions of the diffraction probability from non-relativistic and also classical theory, in the hope to acquire more intuition for the dynamics of the process of interest. We start considering the non-relativistic quantum description in terms of the Schr\"{o}dinger equation. When coupled to the vector potential only, the single component Hamiltonian of the Schr\"{o}dinger equation reads
\begin{equation}
	H = \frac{1 }{2m}\left(\vec p -\frac{e}{c}\vec A\right)^2\,,
	\label{eq:hamiltonian_pauli}
\end{equation}
with the time-evolution implied by Eq. \eqref{eq:dirac_equation}. As in the relativistic case, we split the Hamiltonian \eqref{eq:hamiltonian_pauli} again into a kinetic term $H_0$ and an interaction term $V$ by
\begin{subequations}%
\begin{align}%
	H_0 &= \frac{\vec p^2}{2m}\label{eq_pauli_free_hamiltonian}\\
	V&=\frac{e^2\vec{A}^2}{2mc^2} - \frac{e\vec{A}\cdot \vec p}{mc}\,,\label{eq_schroedinger_potential}
\end{align}%
\end{subequations}%
where the contribution $\vec p \cdot \vec A$ with $\vec p$ acting at $\vec A$ is zero for the assumed vector potential \eqref{eq:vector_potential}.

In the non-relativistic case, the frequencies $\omega_n = \mathcal{E}_n/\hbar$ with momenta $\vec p_n=[(n-1)\hbar k_k,0,p_3]^T$ are now given by the energy eigenvalues of \eqref{eq_pauli_free_hamiltonian} as
\begin{equation}
 \mathcal{E}_n=\frac{(n-1)^2 \hbar^2 k_L^2}{2m} + \frac{p_3^2}{2 m}\,.\label{eq:pauli_eigen_energies}
\end{equation}
Analogously, we expand the wave function $\Psi(\vec x ,t)$ as
\begin{equation}
	\Psi(\vec x ,t) =\sum_{n}c_n(t)\psi_n(\vec x)\,,
\end{equation}
with the single component plane waves
\begin{equation}
	\psi_n(\vec x )=\sqrt{\frac{k_L}{2\pi}} e^{i\vec p_n \cdot \vec x/\hbar}\,.\label{eq:bi_spinor_plane_waves_schrodinger}
\end{equation}

\subsection{\label{sec:perturbative_solution_s}Perturbative solution}

A system of differential equations for the Schrödinger Hamiltonian \eqref{eq:hamiltonian_pauli} with the expansion coefficients $c_n(t)$ similarly to \eqref{eq:inserted_dirac_equation} can be established \cite{ahrens_2012_phdthesis_KDE}, with the interaction operator matrix elements in discrete momentum space
\begin{subequations}\label{eq:matrix_interaction_schrodinger_equation}%
\begin{align}%
&V_{n,n^\prime} =\frac{1}{\lambda}\int_{-\lambda/2}^{\lambda/2} d\vec x\, \psi_n^*(\vec x ) \left[\frac{e^2\vec{A}^2}{2mc^2} - \frac{e\vec{A}\cdot \vec p}{mc}\right]\psi_{n^\prime}(\vec x )\nonumber\\
&= \frac{e^2A_0^2\sin^2(\omega t)\xi^2(t)}{8mc^2}(\delta_{n,n^\prime-2}+2\delta_{n,n^\prime}+\delta_{n,n^\prime+2})\label{eq:matrix_interaction_schrodinger_equation_ponderomotive}\\
&\,\quad-\frac{ e A_0 p_3 \sin(\omega t)\xi(t)}{2mc}(\delta_{n,n^\prime-1}+\delta_{n,n^\prime+1})\,.\label{eq:matrix_interaction_schrodinger_equation_pA}
\end{align}%
\end{subequations}%
We are interested in using the interaction \eqref{eq:matrix_interaction_schrodinger_equation} for obtaining the perturbative solution of $U_{2,0}$ for the case of the Schr\"{o}dinger equation. In contrast do the relativistic quantum theory with the Dirac equation also a second next neighbor coupling arises in Eq. \eqref{eq:matrix_interaction_schrodinger_equation_ponderomotive}, where the vector potential enters the Hamiltonian quadratically. This implies non-vanishing contributions also from first order time-dependent perturbation theory \cite{sakurai2014modern}
\begin{equation}
  U^\textrm{st}(t,0)=\left(-\frac{i}{\hbar}\right)\int^t_0 \,dt^\prime \tilde V(t^{\prime})\,,\label{eq:general_first_perturbation_theory}
\end{equation}
which can be written as
\begin{equation}
  U_{2,0}^\textrm{st}(t,0)
  = \left(-\frac{i}{\hbar}\right)\int_{0}^{t}\,dt^\prime e^{i(\omega_2-\omega_{0})t^\prime} V_{2,0}(t^\prime)\label{eq:first_order_perturbation_schrodinger}
\end{equation}
in the component notion of the interaction picture. As for the relativistic case, we have $\omega_2=\omega_0$ and with setting $\xi(t)=1$, the only relevant matrix element in the interaction \eqref{eq:matrix_interaction_schrodinger_equation} results in
\begin{equation}
 U_{2,0}^\textrm{st}(t,0)
  = \left(-\frac{i}{\hbar}\right) \left( \frac{e^2 A_0^2}{8mc^2} \right)\int_{0}^{t}\,dt^\prime \sin^2(\omega t')\,.
\end{equation}
The $\sin^2$ integral evaluates to $t/2$ for integration times equal to multiples of a half laser period $t=n \pi/\omega$, $n \in \mathbb{N}$. Since the oscillatory part of the integral gets negligibly small as compared to the integral value for large times $t$, we can approximate
\begin{equation}
 U_{2,0}^\textrm{st}(t,0) \approx -i \frac{e^2 A_0^2t}{16mc^2 \hbar}\,.\label{eq:first_order_non_relativistic_perturbation}
\end{equation}
The next-neighbor coupling in \eqref{eq:matrix_interaction_schrodinger_equation_pA} also implies a lowest order non-vanishing contribution for $U_{2,0}$ in second order time-dependent perturbation theory
\begin{multline}
  U_{2,0}^\textrm{nd}(t,t_0) = \left(-\frac{i}{\hbar}\right)^2\int_{t_0}^t \,dt^\prime \int_{t_0}^{t^\prime}dt^{\prime\prime}\\
  \times e^{i(\omega_{2}-\omega_{1}) t^{\prime}}V_{2,1}(t^\prime) e^{i(\omega_{1}-\omega_{0})t^{\prime\prime}}V_{1,0}(t^{\prime\prime})\,.\label{eq:second_order_perturbation_schrodinger}
\end{multline}
Setting again $\xi(t)=1$ results in
\begin{equation}
  U_{2,0}^\textrm{nd}(t,0) = - \left(\frac{e A_0 p_3}{4mc\hbar}\right)^2 \Xi^+\,,\label{eq:pauli_second_perturbation_expanded}
\end{equation}
where we reuse expression \eqref{eq:dirac_perturbation_time_integral} for the time integral. However, with the non-relativistic energies \eqref{eq:pauli_eigen_energies}, the denominator in \eqref{eq:two_coefficients_l_photons} turns into
\begin{equation}%
  \omega_1 - \omega_0 + a \omega = - \frac{\hbar k_L^2}{2 m} + a c k_L = - \frac{k_L}{2 m}(\hbar k_L - 2 a m c)\,,
\end{equation}%
\noindent such that the sum \eqref{eq:two_coefficients_l_sum} can be approximated as
\begin{multline}
 F^+=- \frac{2 m}{k_L}\left(\frac{1}{\hbar k_L - 2 m c}+\frac{1}{\hbar k_L + 2 m c}\right)\\
 = - \frac{2 m}{k_L}\frac{2 \hbar k_L}{\hbar^2 k_L^2 - 4 m^2 c^2}
 \approx \frac{\hbar}{m c^2}\,,
\end{multline}
where we assume $\hbar k_L \ll m c$. Inserting $-i F^+ t$ as solution of $\Xi^+$ into Eq. \eqref{eq:pauli_second_perturbation_expanded} results in
\begin{equation}%
  U_{2,0}^\textrm{nd}(t,0) = i \frac{e^2 A_0^2 p_3^2 t}{16 m^3 c^4 \hbar}\,.\label{eq:second_order_non_relativistic_perturbation}
\end{equation}%
\noindent Finally, the sum of first and second order perturbation in non-relativistic quantum theory can be written as
\begin{multline}
 U_{2,0}(t,0) = U_{2,0}^\textrm{st}(t,0) + U_{2,0}^\textrm{nd}(t,0)\\
 \approx -i \frac{e^2 A_0^2t}{16mc^2 \hbar} \left( 1 - \frac{p_3^2}{m^2 c^2} \right)\,.\label{eq:non_relativistic_propagator}
\end{multline}
The result \eqref{eq:non_relativistic_propagator} reveals a specific velocity dependence of the diffraction amplitude that differs from the standard ponderomotive approximation of the Kapitza-Dirac effect \cite{batelaan_2000_KDE_first,batelaan_2007_RMP_KDE}. The $p_3^2/(m^2c^2)$ correction arises from the $\vec A \cdot \vec p$ term in the interaction Hamiltonian \eqref{eq_schroedinger_potential} and is essential for the identification with the relativistic quantum result \eqref{eq:dirac_perturbation_approximation}, despite a modified functional dependence of $p_3$ in the context of non-relativistic theory. We have obtained this velocity-dependent correction to the Kapitza-Dirac diffraction amplitude by extending the perturbative calculation with including a second order term \eqref{eq:second_order_perturbation_schrodinger}-\eqref{eq:second_order_non_relativistic_perturbation}, for explicitly representing the transverse momentum dependence.

\section{\label{procedure_poderomotive_tkde}Non-relativistic classical description}

The classical motion of an electron in a standing light wave leads to an effective ponderomotive potential, which can be directly related to the electron diffraction process in the Kapitza-Dirac effect \cite{batelaan_2000_KDE_first,batelaan_2007_RMP_KDE,ivanov_batelaan_2004_moving_grating_KDE}, where we follow and extend the calculation in references \cite{batelaan_2000_KDE_first,batelaan_2007_RMP_KDE} in the following. We begin the derivation of the ponderomotive potential with the Lorentz force
\begin{equation}
 \vec F = e \left[ \vec E(\vec x,t) + \frac{\vec p}{m c} \times \vec B(\vec x,t) \right] \,, \label{eq:lorentz-force}
\end{equation}
which the electron as classical, point-like particle is subject to. The electric and magnetic fields in the Lorentz force are implied by the vector potential \eqref{eq:vector_potential}, through the relations \cite{jackson_1999_classical_electro_dyamics}
\begin{subequations}\label{eq:em_fields}%
	\begin{align}%
		\vec E(\vec x,t) &= - \frac{1}{c}\frac{\partial \vec A}{\partial t}
		&=& -A_0 k_L\cos( k_Lx)\cos(\omega t) \vec e_3 \label{eq:electric_amplitude}\\
		\vec B(\vec x,t) &= \vec \nabla \times \vec A
		&=& A_0 k_L \sin(k_Lx)\sin(\omega t) \vec e_2\,,\label{eq:magnetic_amplitude}
	\end{align}%
\end{subequations}%
for the situation with $\xi(t)=1$. For an initial particle momentum $\vec p=p_3\vec e_3$ and the corresponding initial position trajectory $\vec x=(x,y,z+p_3 t/m)$ this yields
\begin{multline}
 \vec F = - e A_0 k_L \cos(k_L x) \cos(\omega t) \vec e_3
 \\- \frac{e A_0 k_L p_3}{m c} \sin(k_L x) \sin(\omega t) \vec e_1 \,. \label{eq:ponderomotive-zero-order-force-nonaveraged}
\end{multline}
The Lorentz force \eqref{eq:lorentz-force} implies the modified particle momentum
\begin{multline}
 \tilde {\vec p} = \int \vec F(t') dt'
 = \left[- \frac{e A_0}{c} \cos(k_L x) \sin(\omega t) + p_3 \right]\vec e_3 \\
 + \frac{e A_0 p_3}{m c^2} \sin(k_L x) \cos(\omega t) \vec e_1 \label{eq:ponderomotive-first-oder-velocity}
\end{multline}
and consequently also the modified particle position
\begin{multline}
 \tilde {\vec x} = \int \frac{\tilde {\vec p}(t')}{m} dt'
 = \vec x + \frac{e A_0}{k_L m c^2} \cos(k_L x) \cos(\omega t) \vec e_3\\
 + \frac{e A_0 p_3}{k_L m^2 c^3} \sin(k_L x) \sin(\omega t) \vec e_1 \,. \label{eq:ponderomotive-first-oder-position}
\end{multline}
Here, we assume that except the original particle position $\vec x$ and momentum $\vec p$ no further additional integration constants arise when computing the anti-derivatives, for the case that the electron was adiabatically entering the laser field in the distant past. The oscillations
\begin{equation}
 \delta \vec x = \tilde {\vec x} - \vec x\,,\qquad \delta \vec p = \tilde {\vec p} - \vec p\,,
\end{equation}
are now emerging in Eqs. \eqref{eq:ponderomotive-first-oder-velocity} and \eqref{eq:ponderomotive-first-oder-position}, if compared to the original position $\vec x$ and momentum $\vec p$. At these updated positions, the electro-magnetic fields can be approximated as
\begin{subequations}%
\begin{align}%
 \tilde {\vec E} &= \vec E(\tilde {\vec x},t) \approx \vec E(\vec x,t) + \delta \vec E\\
 \tilde {\vec B} &= \vec B(\tilde {\vec x},t) \approx \vec B(\vec x,t) + \delta \vec B\,.
\end{align}%
\end{subequations}%
The small quantities $\delta \vec E$ and $\delta \vec B$ are taken as the terms of a first order Taylor expansion of the electro-magnetic field, which contains the field derivatives
\begin{subequations}\label{eq:partial_E_B}%
\begin{align}%
\frac{\partial}{\partial x} \vec E(\vec x,t)&= A_0k_L^2\sin(k_L x)\cos(\omega t) \vec e_3\\
\frac{\partial}{\partial x} \vec B(\vec x,t)&= A_0k_L^2\cos(k_L x)\sin(\omega t) \vec e_2\,.
\end{align}%
\end{subequations}%
We can thus combine
\begin{subequations}%
\begin{align}%
 \delta \vec E =& \delta x \frac{\partial}{\partial x} \vec E = \frac{e A_0^2 k_L p_3}{m^2 c^3} \sin^2(k_L x) \sin(\omega t) \cos(\omega t) \vec e_3 \,,\label{eq:ponderomotive-electric-field-first-order}\\
 \delta \vec B =& \delta x \frac{\partial}{\partial x} \vec B = \frac{e A_0^2 k_L p_3}{m^2 c^3} \sin(k_L x) \cos(k_L x) \sin^2(\omega t) \vec e_2 \,.\label{ponderomotive-magnetic-field-first-order}
\end{align}%
\end{subequations}%
Therefore, the fields $\tilde {\vec E}$ and $\tilde {\vec B}$ at the updated position $\tilde {\vec x}$ and momentum $\tilde {\vec p}$ result in the modified force
\begin{equation}
 \tilde {\vec F} \approx e \left[\vec E + \delta \vec E + \frac{\vec p + \delta \vec p}{m c} \times (\vec B + \delta \vec B) \right]\,.\label{eq:ponderomotive-first-order-force-nonaveraged}
\end{equation}
The ponderomotive force, which we want to discuss here, is associated with the average of the force \eqref{eq:ponderomotive-first-order-force-nonaveraged} over one laser cycle
\begin{equation}
 \braket{\tilde {\vec F}} = \frac{\omega}{2 \pi} \int_{t_0}^{t_0+\frac{2 \pi}{\omega}} \tilde {\vec F} dt \,.
\end{equation}
We see that the cycle averages for the terms $\braket{\vec E}$, $\braket{\delta \vec E}$ and $\braket{\vec p \times \vec B}$ in the expansion of \eqref{eq:ponderomotive-first-order-force-nonaveraged} are zero. The contribution $\braket{\delta \vec p \times \delta \vec B}$ is the only term which scales with the third power of the small quantity $A_0$, and we neglect it for this reason. It remains
\begin{subequations}\label{eq:ponderomotive_force_all}%
\begin{align}%
 &\braket{\tilde {\vec F}} \approx \frac{\omega}{2 \pi} \int_{t_0}^{t_0+\frac{2 \pi}{\omega}} \frac{e}{mc} (\delta \vec p \times \vec B + \vec p \times \delta \vec B) dt \label{eq:ponderomotive-force_contributions}\\
 &=\frac{k_Le^2 A_0^2}{m c^2}\left( 1  - \frac{p_3^2}{m^2 c^2} \right) \sin(k_L x) \cos(k_L x) \vec e_1 \\
 &\qquad \qquad \qquad \qquad \qquad \times \frac{\omega}{2 \pi} \int_{t_0}^{t_0 + \frac{2 \pi}{\omega}} \sin^2(\omega t) dt\\
 &=\frac{k_Le^2 A_0^2}{2 m c^2}\left( 1  - \frac{p_3^2}{m^2 c^2} \right) \sin(k_L x) \cos(k_L x) \vec e_1\,,\label{eq:ponderomotive-force}
\end{align}%
\end{subequations}%
where the cycle average of the transverse component of $\braket{\delta\vec p \times \vec B}$ is zero as well.

The ponderomotive force expression \eqref{eq:ponderomotive-force} has some similarities with the non-relativistic perturbative solution \eqref{eq:non_relativistic_propagator} of the Schr\"{o}dinger equation. Consequently, and consistent with the considerations in reference \cite{ivanov_batelaan_2004_moving_grating_KDE}, we show in the following that \eqref{eq:ponderomotive-force} actually causes the same quantum state propagation expression \eqref{eq:non_relativistic_propagator}, by computing the ponderomotive potential from the ponderomotive force, and then calculating the time-evolution which is implied by the potential.

The ponderomotive potential is obtained by integrating the ponderomotive force along the direction of the only non-trivial spacial dependence (ie. along the laser propagation direction), which yields
\begin{multline}
	V_{\textrm{pond}} = - \int \braket{\tilde{\vec F}} \cdot \vec e_1 dx\\
  = \frac{e^2 A_0^2}{4 m c^2}\left( 1  - \frac{p_3^2}{m^2 c^2} \right) \cos^2(k_L x)\,. \label{eq:unshifted-ponderomotive-potential}
\end{multline}
Computing the momentum space matrix elements of this ponderomotive potential similarly to the procedure in Eq. \eqref{eq:matrix_interaction_schrodinger_equation_ponderomotive} results in the expressions
\begin{align}%
&V^\textrm{pond}_{n,n^\prime} =\frac{1}{\lambda}\int_{-\lambda/2}^{\lambda/2} d\vec x\, \psi_n^*(\vec x ) V_\textrm{pond}\psi_{n^\prime}(\vec x ) \label{eq:ponderomotive_potential}\\
&= \frac{e^2 A_0^2}{16 m c^2}\left( 1  - \frac{p_3^2}{m^2 c^2} \right)(\delta_{n,n^\prime-2}+2\delta_{n,n^\prime}+\delta_{n,n^\prime+2})\,.\nonumber
\end{align}%
The first order time-dependent perturbation theory expression of Eq. \eqref{eq:ponderomotive_potential} then yields
\begin{multline}
  U_{2,0}^\textrm{st}(t,0)
  = \left(-\frac{i}{\hbar}\right)\int_{0}^{t}\,dt^\prime e^{i(\omega_2-\omega_{0})t^\prime} V^{\textrm{pond}}_{2,0}\\
 = -i \frac{e^2 A_0^2t}{16mc^2 \hbar} \left( 1 - \frac{p_3^2}{m^2 c^2} \right)\,,\label{eq:non_relativistic_classical_propagator}
\end{multline}
which is indeed consistent with the result \eqref{eq:non_relativistic_propagator} of the non-relativistic perturbative quantum calculation in section \ref{sec:perturbative_solution_s}.

\section{\label{procedure_relativistic_poderomotive_tkde}Relativistic classical description}

The non-relativistic results \eqref{eq:non_relativistic_propagator} and \eqref{eq:non_relativistic_classical_propagator} for the electron dynamics are not consistent with the relativistic quantum solution in \eqref{eq:dirac_perturbation_approximation}, which is a reason to investigate the relativistic classical motion of the electron as well. While various forms of the relativistic ponderomotive potential have been discussed in the literature \cite{bauer_1995_ponderomotive_chaos,Kaplan_2005_ponderomotive_ultra_intense,Pokrovsky_2005_ponderomotive_reversal,Smorenburg_2011_polarization_dependent_ponderomotive,Ribbing_2025_general_ponderomotive}, the specific derivation for the standing-wave geometry of the Kapitza-Dirac effect, especially concerning the electron's transverse momentum dependence, which is of interest here, requires a careful treatment of the calculation. We thus extend the non-relativistic ponderomotive calculation from Batelaan et al. \cite{batelaan_2000_KDE_first,batelaan_2007_RMP_KDE}, which is formulated in the specific conventions of the Kapitza-Dirac effect, into the framework of the relativistic equations of motion, introduced below, in Eqs. \eqref{eq:relativistic_motion_equation}. This extension allows us to identify specific relativistic corrections---in particular the factor of $5/2$ in Eq.~\eqref{eq:relativistic_ponderomotive_force}---which modifies the non-relativistic result and enables a direct comparison with the relativistic quantum calculation. Thus, our aim is, in the following, to discuss where corrections of the relativistic equations of motion occur in the non-relativistic limit of the calculation.

The classical motion of an electron in a relativistic framework \cite{jackson_1999_classical_electro_dyamics} can be denoted in the lab-frame by the coupled differential equations
\begin{subequations}
\label{eq:relativistic_motion_equation}
\begin{align}
\frac{d\vec{p}}{dt} &= \vec F = e \left[ \vec{E}(\vec{x},t) + \frac{\vec{p}}{\gamma(\vec{p}) m c} \times \vec{B}(\vec{x},t) \right]\,, \label{eq:relativistic_lorentz-force}\\
\frac{d\vec{x}}{dt} &= \frac{\vec{p}}{\gamma(\vec{p}) m}\,. \label{eq:relativistic_displacement}
\end{align}
\end{subequations}
Eq. \eqref{eq:relativistic_lorentz-force} originates from a covariant formulation of the Lorentz force and generalizes the Lorentz force \eqref{eq:lorentz-force}. The right-hand side of Eq. \eqref{eq:relativistic_displacement} originates from the four-velocity and four-mometum, with $\vec p = m \gamma \vec v$ with respect to the lab-time $t$. The relativistic gamma factor
\begin{equation}
\gamma(\vec{p}) = \sqrt{1 + \frac{\vec{p}^{\,2}}{m^2c^2}}
\end{equation}
is implied from \eqref{eq:relativistic_energy_momentum_relation} and $E=\gamma m c^2$. We first solve the dynamics of the classical relativistic electron trajectory numerically, according of Eqs. \eqref{eq:relativistic_motion_equation} for parameters, at which the non-linear contributions to the electron dynamics appear to be small. Suitable parameters for the potential \eqref{eq:vector_potential} with envelope \eqref{eq:temporal_envelope_function} are $q A_0 = 10^{-3} \, m c^2$, $\hbar k_L = 0.02 \,m c$, $T=150 \times 2 \pi/\omega$ and $\Delta T=5 \times 2 \pi/\omega$. The initial particle position is $x=\lambda/8$, which is at a location of the periodic laser field, where we know from the non-relativistic calculation that the ponderomotive deflection is largest. We further set the initial particle parameters $y=z=0$ and $\vec p=p_3 \vec e_3$ for the numeric simulation. The result of the numeric calculation is displayed in Fig. \ref{fig:Simulation_and_Predicted} as green dotted line, where we know from the calculation in Eqs. \eqref{eq:ponderomotive_force_all}-\eqref{eq:non_relativistic_classical_propagator} that the ponderomotive amplitude (which is proportional to $p_1$) converts into a diffraction amplitude. The diffraction amplitude, in turn, converts further into a diffraction probability by taking the absolute value square of the wave function, or equivalently its propagation equation \eqref{eq:pertubation_theory}, which is then proportional to $p_1^2$. Therefore, we display $p_1^2$ at the ending time $T$ of the numerical particle trajectory simulation in Fig. \ref{fig:Simulation_and_Predicted}, and find proper agreement with the relativistic quantum solutions. In other words, we find a correspondence between the relativistic quantum and classical results, which is established by a matching of the transverse momentum dependence. We will substantiate this quantitative agreement by comparing the perturbative descriptions in the following, resulting in a factor 5/2 in Eq. \eqref{eq:relativistic_ponderomotive_force} for the relativistic classical calculation. This agrees with the coefficient in the relativistic quantum solution \eqref{eq:dirac_perturbation_approximation} and provides us another analytic cross-check for confirming the agreement between relativistic quantum and relativistic classical theory, rather than relying solely on a numerical comparison.

With having the evidence, that the classical equations of motion \eqref{eq:relativistic_motion_equation} appear to be consistent with the quantum solutions, we would like to point out where the non-relativistic limit of the dynamics leads to relativistic corrections in the non-relativistic dynamics as described in the previous section. Details of this calculation appear in the more verbose calculation description in the supplemental material \cite{supplement_material_real_ponder}. The difference of the relativistic equations of motion \eqref{eq:relativistic_motion_equation} as compared to the classical ones in Eqs. \eqref{eq:lorentz-force}, \eqref{eq:ponderomotive-first-oder-velocity} and \eqref{eq:ponderomotive-first-oder-position} is the inverse gamma factor
\begin{align}
\gamma^{-1}(\tilde{\vec{p}})
&\approx 1 - \frac{\tilde{p}_1^2}{2m^2c^2} - \frac{\tilde{p}_3^2}{2m^2c^2} +\frac{3\tilde p_1^2\tilde p_3^2}{4m^4c^4}. \label{eq:gamma_factor}
\end{align}
For the motion in first order \eqref{eq:ponderomotive-first-oder-velocity} the gamma factor evaluates as
\begin{multline}
\hspace{-0.3 cm}\gamma^{-1}(\tilde{\vec{p}})
\approx 1 - \frac{e^2A_0^2p_3^2}{2m^4c^6}\sin^2(k_Lx)\cos^2(\omega t) - \frac{p_3^2}{2m^2c^2}\\
- \frac{e^2A_0^2}{2m^2c^4}\cos^2(k_Lx)\sin^2(\omega t)
+ \frac{eA_0p_3}{m^2c^3}\cos(k_Lx)\sin(\omega t)\,, \label{eq:inverse_gamma_factor}
\end{multline}
where we neglect terms of order $\mathcal{O}(p_3^3)$ and $\mathcal{O}(A_0^3)$ and higher. In what follows, the updated $x$-component of the particle position $\tilde {\vec x}$ is of relevance, as this determines the $x$-dependent variation of the electro-magnetic field in the derivation of the ponderomotive force. Thus, by integrating the particle position according to Eq.~\eqref{eq:relativistic_displacement}, we obtain the expression
\begin{subequations}%
\begin{align}%
\delta x &= \frac{eA_0p_3}{k_L m^2c^3}\sin(k_Lx)\sin(\omega t) \\
&\quad + \frac{e^2A_0^2p_3^2}{2 k_L m^4c^6}\cos(k_Lx)\sin(k_Lx)\sin^2(\omega t)\,, \label{eq:oscillation_position_x}
\end{align}%
\end{subequations}%
which is to be matched and compared with the $x$-component of the non-relativistic expression \eqref{eq:ponderomotive-first-oder-position}. The new term in \eqref{eq:oscillation_position_x} still needs to be multiplied with the electro-magnetic fields in subsequent calculation steps and will result in terms $\mathcal{O}(A_0^3)$, which we neglect. Therefore, the expression \eqref{eq:oscillation_position_x} which originates from the expansion \eqref{eq:gamma_factor} of the relativistic gamma factor $\gamma^{-1}(\tilde{\vec{p}})$ does not give any extra contributions for the ponderomotive force.

Additionally to the relativistic correction of the particle position $\vec x$, The inverse gamma factor introduces a second correction, which appears in the $\tilde{\vec{p}}\gamma(\tilde{\vec{p}})^{-1}$ expression of the Lorentz force \eqref{eq:relativistic_lorentz-force}
\begin{align}
\frac{\tilde{\vec{p}}}{\gamma(\tilde{\vec{p}})}
&=\mathcal{A}+\mathcal{B}+\mathcal{C}+\mathcal{D}+\mathcal{F}+\mathcal{G} + \mathcal{O}(p_3^3) + \mathcal{O}(A_0^3)\,,
 \label{eq:updated_momentum_with_gamma_minus}
\end{align}
with terms
\begin{subequations}
	\begin{align}
		\mathcal{A} &= \frac{eA_0p_3}{mc^2}\sin(k_Lx)\cos(\omega t) \vec{e}_1\,, \\
		\mathcal{B} &= \frac{e^2A_0^2p_3^2}{m^3c^5}\sin(k_Lx)\cos(k_Lx)\sin(\omega t)\cos(\omega t) \vec{e}_1\,, \\
		\mathcal{C} &= p_3 \vec{e}_3\,, \\
		\mathcal{D} &= -\frac{3e^2A_0^2p_3}{2m^2c^4}\cos^2(k_Lx)\sin^2(\omega t) \vec{e}_3\,, \\
		\mathcal{F} &= \frac{3eA_0p_3^2}{2m^2c^3}\cos(k_Lx)\sin(\omega t) \vec{e}_3\,, \\
		\mathcal{G} &= -\frac{eA_0}{c}\cos(k_Lx)\sin(\omega t) \vec{e}_3\,. \label{eq:some_terms}
	\end{align}
\end{subequations}
The relevant contributions for the time-averaged ponderomotive force of Eq. \eqref{eq:relativistic_lorentz-force} consist of the three expressions
\begin{multline}\label{eq:relativistic_ponderomotive_force}
\hspace{-0.4 cm}\braket{\tilde{\vec{F}}}
= \frac{e}{mc} \left( \braket{\mathcal{F} \times \vec{B}} + \braket{\mathcal{G} \times \vec{B}} + \braket{\mathcal{C} \times \delta \vec{B}} \right) \\
= \left[ \frac{e^2A_0^2k_L}{2mc^2} \left(1 - \frac{5}{2}\frac{p_3^2}{m^2c^2}\right) \sin(k_Lx)\cos(k_Lx) \right] \vec{e}_1\,,
\end{multline}
after eliminating vanishing terms and omitting higher-order powers of $p_3$ and $A_0$, as discussed in the comprehensive derivation of the ponderomotive potential in the supplemental material \cite{supplement_material_real_ponder}. Following the steps \eqref{eq:unshifted-ponderomotive-potential}-\eqref{eq:non_relativistic_classical_propagator} yields the same type of matrix element for the time-evolution operator $U_{2,0}(t,0)$ as in the relativistic quantum calculation \eqref{eq:dirac_perturbation_approximation}, but now with the correct factor $5/2$ in front of the reduced momentum $[p_3/(mc)]^2$. A comparison with \eqref{eq:ponderomotive_force_all} reveals that $\braket{\mathcal{F} \times \vec{B}}$ is the new contribution from the relativistic classical calculation, which provides the matching relativistic correction to the non-relativistic result and stems from the gamma factor \eqref{eq:inverse_gamma_factor}.

\section{\label{sec:discussion}Discussion}

The calculations presented above establish specific connections between the quantum and classical descriptions of the Kapitza-Dirac effect. The velocity-dependent correction in the Schr\"odinger equation \eqref{eq:non_relativistic_propagator}, the extension of the ponderomotive force \eqref{eq:ponderomotive_force_all} to the relativistic regime \eqref{eq:relativistic_ponderomotive_force}, and the quantitative matching between the relativistic quantum and classical results \eqref{eq:dirac_perturbation_approximation} and \eqref{eq:relativistic_ponderomotive_force} provide a consistent framework for understanding the diffraction process across different theoretical descriptions. However, only the relativistic quantum calculation in section \ref{sec:energy_eigenstate_dynamics} with the Dirac equation allows for the investigation of negative intermediate states in comparison to positive intermediate states. Interesting is their different interpretation circumstance: The situation of positive intermediate states in the contribution \eqref{eq:propagator_coupling_positive_up} could be matched to a classical interpretation of motion. In contrast, the contribution \eqref{eq:propagator_coupling_negative_up} is putting an additional obstacle on the attempt of an intuitive understanding of the process, since \eqref{eq:propagator_coupling_negative_up} originates from the coupling to negative electron states, which does not exist in classical theory. We further observe that the transverse velocity dependence of the diffraction amplitude \eqref{eq:dirac_perturbation_approximation} in relativistic quantum theory is contributed by the $s^3_{n,n'}$ term \eqref{eq:propagator_coupling_positive_up} of positive intermediate state coupling, which can be associated with the $\vec A \cdot \vec p$ contribution of the dynamics from the Schr\"{o}dinger equation in non-relativistic quantum theory. This $\vec A \cdot \vec p$ contribution, however, matches the $\vec p \times \delta \vec B$ contribution of the classical motion in the non-relativistic description and the $\braket{\mathcal{C}\times \vec B}$ contribution in the relativistic description. One may thus identify the positive intermediate state coupling in the relativistic quantum theory with the classical motion of a deflected electron trajectory component along the laser propagation direction, which causes the variation of the magnetic field $\delta \vec B$.

Still, it is \emph{only} the negative state coupling \eqref{eq:propagator_coupling_negative_up}, which contributes an expression in the diffraction amplitude \eqref{eq:dirac_perturbation_approximation}, which is partially independent of the transverse velocity $p_3$! This negative state coupling contribution can be associated with the $\vec A^2$ term in the Schr\"{o}dinger equation, which in turn matches the resulting diffraction probability component of the $\delta \vec p \times \vec B$ motion of the ponderomotive potential in the non-relativistic case and $\braket{(\mathcal{F}+\mathcal{G})\times \vec B}$ in the relativistic case. We emphasize that the term $\delta \vec p \times \vec B$ (or $\braket{(\mathcal{F}+\mathcal{G})\times \vec B}$, respectively) corresponds to the `conventional' classical motion component of the electron along the electric field, which causes the ponderomotive potential from the Lorentz force, as described by Batelaan \cite{batelaan_2000_KDE_first,batelaan_2007_RMP_KDE}.

At this point, we mention that the dynamics of the Kapitza-Dirac effect appears symmetric for the case of electrons (positive states) and positrons (negative states), since the charge conjugation operation $\Psi_c = \hat C \bar{\Psi}^T$ is equivalent so a sign flip of the elementary charge $e$ or, correspondingly, the vector potential $A_0$ in the Dirac equation \cite{Greiner_1985_strong_field_QED}.

\begin{figure}%
	\includegraphics[width=0.42 \textwidth]{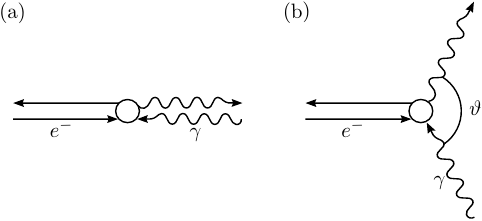}
	\caption{\label{fig:COM_interaction_layout} Interaction geometry of Compton scattering, in a frame of reference in which the electron is reversing its momentum $\vec p$. (a) In this reference frame the photon is back-scattering by $180^\circ$ for a situation of the Kapitza-Dirac effect with zero transverse momentum $p_3=0$. (b) In contrast, the photon scatters off by some angle $\vartheta$ for situations with large transverse momenta $p_3 \gg m c$ in the Kapitza-Dirac effect.}
\end{figure}

In order to summarize the similarities of the terms in non-relativistic and relativistic quantum- and classical theory, we need to distinguish between the situation of small transverse momenta ($p_3\ll mc$) along the laser polarization direction and the situation of relativistic transverse momenta ($p_3\gg mc$) in the Kapitza-Dirac effect. For a descriptive interpretation we want to consider a frame of reference, where the electron is reversing its momentum $\vec p$ in the context of an analogous Compton scattering process. In such an electron centered frame, the photon performs a head-on collision for the situation with $p_3=0$, as sketched in Fig. \ref{fig:COM_interaction_layout}(a). However, the photon scatters off the electron at a larger angle $\vartheta$ for the situation with $p_3 \gg m c$, as illustrated in Fig. \ref{fig:COM_interaction_layout} (b).

\subsection{Small transverse electron momentum $p_3\ll mc$ situation in Fig. \ref{fig:COM_interaction_layout}(a)\label{sec:small_transverse_situation}}

We can make a series of statements about properties of the investigated negative state coupling for the two distinct scenarios in Fig. \ref{fig:COM_interaction_layout}, for which in Fig. \ref{fig:COM_interaction_layout}(a) ie. $p_3\ll mc$ the following properties can be associated with each other:
\begin{enumerate}
\item The coupling to negative intermediate states [red dashed line in Fig. \ref{fig:constituents_two_photon} and \eqref{eq:propagator_coupling_negative_up} in the low transverse momentum approximation] in the electron-light interaction of the Kapitza-Dirac effect dominates over the coupling to positive intermediate states [blue dash-dotted line and \eqref{eq:propagator_coupling_positive_up}].
\item A virtual electron-position pair [corresponding to $\ket{e^-_2 e^+_1 e^-_0}$ and $\ket{e^-_2 e^+_1 e^-_0,\gamma_+\gamma_-}$ in Eq. \eqref{eq:negative_state_coupling_fully_quantized}] dominantly contributes to the scattering dynamics in Compton scattering in the context of a fully quantized theory.
\item The first order time-dependent perturbative expression $\vec A^2$ [$U^{\textrm{st}}$ in Eqs. \eqref{eq:general_first_perturbation_theory}-\eqref{eq:first_order_non_relativistic_perturbation}] in the expansion of $(\vec p - e \vec A/c)^2$ of the Schr\"odinger equation is dominating the diffraction dynamics of the Kapitza-Dirac effect.
\item The Lorentz force of the electron quiver in the oscillating in electric field [$\delta \vec p \times \vec B$ in Eq. \eqref{eq:ponderomotive-force_contributions} (non-relativistic), $\braket{(\mathcal{G}+\mathcal{F})\times \vec B}$ in Eq. \eqref{eq:relativistic_ponderomotive_force} (relativistic)] dominates in classical ponderomotive motion.
\end{enumerate}
Here, we mention that the points 1 and 2 can be shown to be mathematically equivalent \cite{ahrens_2020_two_photon_bragg_scattering}. Regarding point 2, we further state that the development of investigation in this work assumes alignment of the photon polarizations along the $\vec e_3$ axis.

\subsection{Relativistic transverse electron momentum $p_3\gg mc$ situation in Fig. \ref{fig:COM_interaction_layout}(b)}

For clearness, we also formulate the converse of the statements in section \ref{sec:small_transverse_situation}:
\begin{enumerate}
\item The coupling to positive intermediate states [blue dash-dotted in Fig. \ref{fig:constituents_two_photon}] in the electron-light interaction of the Kapitza-Dirac effect dominates over the coupling to negative intermediate states [red dashed line in Fig. \ref{fig:constituents_two_photon} and \eqref{eq:propagator_coupling_negative_up}].
\item A virtual electron-position pair does \emph{not} dominantly contribute to the scattering dynamics in Compton scattering [corresponding to $\ket{e^-_1}$ and $\ket{e^-_1 ,\gamma_+\gamma_-}$ in Eq. \eqref{eq:positive_state_coupling_fully_quantized}].
\item An extrapolation from small to big transverse electron momenta indicates that also the $\vec p \cdot \vec A$ term [$U^{\textrm{nd}}$ in Eqs. \eqref{eq:second_order_perturbation_schrodinger}-\eqref{eq:second_order_non_relativistic_perturbation}] contributes to the Kapitza-Dirac quantum dynamics.
\item An extrapolation from the non-relativistic limit indicates that also the deflection of the electron trajectory $\delta \vec B$ due to the Lorentz force is influencing the Lorentz force [$\vec p \times \delta \vec B$ in Eq. \eqref{eq:ponderomotive-force_contributions} (non-relativistic), $\braket{\mathcal{C}\times \vec B}$ in Eq. \eqref{eq:relativistic_ponderomotive_force} (relativistic)] and contributes to the quantum dynamics.
\end{enumerate}

\section{Conclusions and Outlook\label{sec:conclusion_and_outlook}}

In this article we have demonstrated that quantum dynamics in standing light waves, especially in the two-photon Kapitza-Dirac effect can be dominantly mediated by virtual negative excitations in the context of a single-particle relativistic quantum description of spin 1/2 particles of the Dirac equation. While the relativistic quantum- and classical descriptions result in consistent diffraction amplitudes with respect to the electron's transverse momentum dependency, non-relativistic dynamics deviates due to a lack of the relativistic gamma factor. Nevertheless, it is interesting that the dominant negative state coupling in time-dependent perturbation theory for the case of small transverse electron momenta ($ p_3\ll mc$) can be equivalently associated with coupling to a virtual electron-positron pair in a fully quantized system, in the context of Compton scattering. In this context, it would be interesting to explore the relation of the dynamics to the Feynman-St\"{u}ckelberg interpretation \cite{Greiner_1985_strong_field_QED}. Does it really mean, that the configuration in Fig. \ref{fig:COM_interaction_layout}(a) is an implementation, in which we have realized an electron which dominantly moves to the future to scatter at the first photon then travels back in time to scatter with the second photon, for finally propagating further into the future? Another research question arises from the consideration that negative states with a certain canonical momentum are subject to the negative dispersion of their negative energy-momentum relation. This negative energy-momentum relation causes an opposite motion as compared to their positive counterparts. The question arises: What happens, if one applies an electric field to the oppositely moving and oppositely charged negative states, and how this opposite motion influences the quantum dynamics of, for example the Kapitza-Dirac diffraction, and does it influence the process of virtual excitations to become real?

\begin{acknowledgments}
The work was supported by the National Natural Science Foundation of China
(Grant No. 12535015).
\end{acknowledgments}

\appendix
\section{Nonrelativistic limit of Dirac equation\label{appendix_A}}

It is illustrative for the perception of negative state coupling to be aware about the derivation of the non-relativistic limit of the Dirac equation. The calculation about the non-relativistic limit can be found in standard text books on advanced quantum theory \cite{wachter_2011_relativistic_quantum_mechanics_2011,schwabl_2000_advanced_quantum_mechanics,greiner_2000_relativistic_quantum_mechanics,Greiner_1985_strong_field_QED} and we follow the notion in references \cite{schwabl_2000_advanced_quantum_mechanics,Greiner_1985_strong_field_QED} in the following. First, we introduce the abbreviation
\begin{equation}
	\vec \pi=\vec p -\frac{e}{c}\vec A\,.
\end{equation}
Inserting the two component approach
\begin{align}
\Psi=
	\begin{pmatrix}
		\varphi\\
		\chi
	\end{pmatrix}
	\exp\left(-\frac{imc^2}{\hbar}t\right)
\end{align}
with spinors $\varphi$ and $\chi$ into the Dirac equation \eqref{eq:dirac_equation} and \eqref{eq:hamiltonian_dirac_s} yields
\begin{equation}
	i\hbar \frac{\partial}{\partial t}
	\begin{pmatrix}
		\varphi\\
		\chi
	\end{pmatrix}
	=c
	\begin{pmatrix}
		\vec\sigma\cdot\vec\pi\chi\\
		\vec\sigma\cdot\vec\pi\varphi
	\end{pmatrix}
	-2mc^2
	\begin{pmatrix}
		0\\
		\chi
	\end{pmatrix}\,.\label{eq:substituted_dirac_equation}
\end{equation}
Assuming that $\chi$ is a slowly varying function in time as compared to the electron rest mass oscillation in the complex plane, ie. $\partial \chi/\partial t \ll m c^2/\hbar$ allows for neglecting the time-derivative of $\chi$. Within this approximation, the lower components in Eq. \eqref{eq:substituted_dirac_equation} can be inverted for
\begin{equation}
\chi =\frac{\vec\sigma\cdot\vec\pi}{2mc}\varphi\label{eq:chi_from_phi}
\end{equation}
and be reinserted into the upper components, which yields
\begin{equation}
	i\hbar\frac{\partial \varphi}{\partial t} = \frac{\vec \pi^2}{2m}\varphi + i \frac{\vec \sigma}{2 m} \cdot \vec \pi \times \vec \pi \, \varphi\,.\label{eq:pauli_equation_non_relativistic_limit}
\end{equation}
Neglecting the spin term of the Pauli equation on the right of the right-hand side results in the Schr\"{o}dinger equation with Hamiltonian \eqref{eq:hamiltonian_pauli}. Note, that we only have arrived at the right-hand side because we inserted a coupling between the upper and lower components \eqref{eq:chi_from_phi} in the upper components in \eqref{eq:substituted_dirac_equation}. In other words: The wave function $\varphi$ of the Schr\"{o}dinger equation (or Pauli equation, respectively) of the upper component in Eq. \eqref{eq:substituted_dirac_equation} had to be passed through the lower components of \eqref{eq:substituted_dirac_equation}, to become \eqref{eq:pauli_equation_non_relativistic_limit}.

We also point out that the approximate block-diagonalization of the Dirac equation in context of the Foldy-Wouthuysen transformations \cite{Foldy_Wouthuysen_1950_FW_transformation,schwabl_2000_advanced_quantum_mechanics,wachter_2011_relativistic_quantum_mechanics_2011,greiner_2000_relativistic_quantum_mechanics} involves transitions to the negative spectrum as well. The term $\vec \pi^2$ in the Schr\"{o}dinger equation arises from the squared appearance $\mathcal{O}^2$ of the block-off-diagonal term $\mathcal{O}= c \vec \alpha \cdot \vec \pi$, which is generating the Foldy-Wouthuysen transformation through the generator $S=-i \beta \mathcal{O}/2 m c^2$. The off-diagonal structure of $\mathcal{O}$ due to the off-diagonal alpha matrices $\vec \alpha$ implies the transition to negative states. This also holds for the relativistic correction of the kinetic energy term, which is proportional to $\mathcal{O}^4$.

\section{Quantum dynamics in the weakly non-perturbative regime \label{sec:appendix_weak_perturbative}}

\begin{figure}%
	\includegraphics[width=0.42 \textwidth]{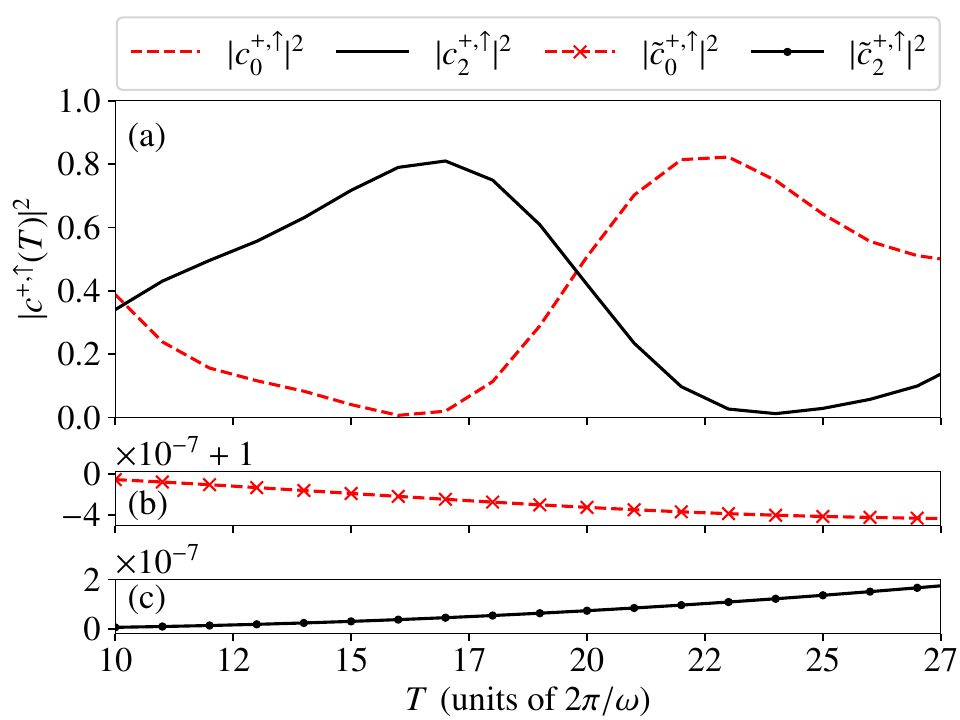}%
	\caption{\label{fig:time_evolution_010}Time evolution of the occupation probabilities in the weakly non-perturbative regime. For the field amplitude we set field amplitude $eA_0 = 0.1mc^2$, with all other parameters being identical to those in Fig.~\ref{fig:time_evolution}. The sinusoidal shape of the Rabi oscillations is distorted in the weakly non-perturbative regime of panel (a), but the significant sensitivity of the quantum dynamics to the coupling to negative states in panels (b) and (c) remains.}
\end{figure}

The perturbative solutions presented the main text are valid for small amplitudes $eA_0/mc^2 \ll 1$ of the interaction potential in time-dependent perturbation theory. In this regime, the population dynamics can be well approximated by sinusoidal Rabi oscillations between the initial and diffracted states, with higher-order channels remaining negligibly populated. However, if one raises the field amplitude $e A_0$, the perturbative treatment becomes increasingly inaccurate and the quantum dynamics enters the weakly non-perturbative regime. In this regime, the characteristic Rabi oscillations become progressively distorted, indicating that higher-order contributions can no longer be neglected and the simple two-level Rabi picture breaks down. We demonstrate this behavior for the case $eA_0 = 0.1mc^2$ in Fig. \ref{fig:time_evolution_010}, where we observe a significant population in higher momentum states ($c_{\pm 3}, c_{\pm 4}$) in our simulation, such that the quantum dynamics of the electron exhibits complex multi-channel interference rather than clean two-level oscillations.

In this context we need to emphasize that our simulation always contains a turn on and turn off period of total duration $2 \Delta T = 10 \times 2\pi/\omega$. Only the plateau phase of the pulse envelope, in which $\xi=1$ in Eq. \eqref{eq:temporal_envelope_function} is varied in Figs. \ref{fig:time_evolution} and \ref{fig:time_evolution_010}. Consequently, the time axis in these figures begins at $T = 10 \times 2\pi/\omega$ rather than $T = 0$. The quantum state, thus, is already evolved from its initial condition in Eq.~\eqref{eq:FSME_initial_conditions} during these periods of ramping the field up and down. Therefore, the ground-state population shown at the left edge of Figs. \ref{fig:time_evolution} and \ref{fig:time_evolution_010} corresponds to $|c_0(20\pi/\omega)|^2$, which is noticeably less than unity for the case of Fig.~\ref{fig:time_evolution_010}. In contrast, the dynamics in Fig.~\ref{fig:time_evolution} is in a regime where perturbation theory is well applicable, and therefore appears to resemble the initial value $|c_0(0)|^2 = 1$.

Regarding the dynamics in Fig.~\ref{fig:time_evolution_010} we emphasize that despite a visible deformation of the Rabi cycles, there is an obvious population transfer between $|c_0^{+,\uparrow}|^2$ and $|c_2^{+,\uparrow}|^2$ in panel (a), which does not alter by toggling off the interactions $V^{\pm,s;\pm,s^\prime}_{n,n^\prime}$. However, the population transfer is smaller by orders of magnitude when switching off $V^{\pm,s;\mp,s^\prime}_{n,n^\prime}$, as shown in panels (b) and (c). This indicates that strong coupling to negative states remains preserved also in the weakly non-perturbative regime.

For the amplitudes of the negative-energy states we observe a non-monotonic dependence on field strength. In a weakly non-perturbative situation the relative contribution of negative-energy intermediate states increases with field amplitude, since the stronger driving field enhances the coupling between positive- and negative-energy sectors. However, when the field becomes stronger and more channels participate, the simple decomposition into negative-energy state contributions becomes less transparent, with higher-order multiphoton processes dominating the dynamics. A complete characterization of the strongly non-perturbative regime, where $eA_0/mc^2 \gtrsim 1$, would require extended numerical investigations, which are beyond the scope of the perturbative analysis presented here, but represent an interesting direction for future work.


\bibliography{bibliography}

\begin{thebibliography}{39}
\bibitem [{sup()}]{supplement_material_real_ponder}%
  \BibitemOpen
  \href@noop {} {\bibinfo {title} {See {S}upplemental {M}aterial at [url will
  be inserted by publisher] for the details of the derivation of the
  relativistic ponderomotive potential.}}\BibitemShut {Stop}%
\end{thebibliography}

\end{document}